\documentclass[fleqn,usenatbib]{mnras}

\usepackage{newtxtext,newtxmath}
\usepackage{makecell}
\usepackage[T1]{fontenc}
\usepackage{ae,aecompl}

\usepackage{graphicx}	% Including figure files
\usepackage{amsmath}	% Advanced maths commands
\usepackage{amsfonts}
\usepackage{hyperref}
\usepackage{pifont}

\usepackage{abbrevs}
\usepackage{xspace}
\usepackage{xcolor}

\newcommand{\orcid}[1]{\href{https://orcid.org/#1}{\includegraphics[width=10pt]{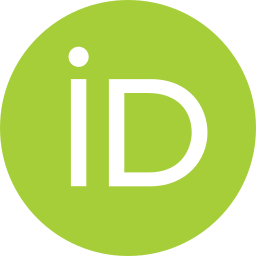}}}

\begingroup
\catcode`\_=\active
\gdef_#1{\ensuremath{\sb{\mathrm{#1}}}}
\endgroup
\mathcode`\_=\string"8000
\catcode`\_=12

\newcommand{\rev}{\textcolor{black}}

\newcommand{\Msolar}{M$_{\odot}$\xspace}

\newcommand{\HI}{H\textsc{i}\xspace}
\newcommand{\HII}{H\textsc{ii}\xspace}
\newcommand{\HeI}{He\textsc{i}\xspace}
\newcommand{\HeII}{He\textsc{ii}\xspace}

\newcommand{\tick}{\hspace{1pt}\ding{51}}
\newcommand{\cross}{\hspace{1pt}\ding{55}}

\newabbrev\ISM{interstellar medium (ISM)}[ISM]
\newabbrev\IGM{Intergalactic Medium (IGM)}[IGM]
\newabbrev\CSM{Circumstellar Medium (CSM)}[CSM]
\newabbrev\WNM{Warm Neutral Medium (WNM)}[WNM]
\newabbrev\WIM{Warm Ionised Medium (WIM)}[WIM]
\newabbrev\CNM{Cold Neutral Medium (CNM)}[CNM]
\newabbrev\IMF{Initial Mass Function (IMF)}[IMF]
\newabbrev\CMF{Core Mass Function (IMF)}[IMF]
\newabbrev\AMR{Adaptive Mesh Refinement (AMR)}[AMR]
\newabbrev\HGB{Horizontal Giant Branch (HGB)}[HGB]
\newabbrev\SFE{Star Formation Efficiency (SFE)}[SFE]
\newabbrev\TSFE{Total Star Formation Efficiency (TSFE)}[TSFE]
\newabbrev\OSFE{Observed Star Formation Efficiency (OSFE)}[OSFE]
\newabbrev\SFR{Star Formation Rate (SFR)}[SFR]
\newabbrev\YSOs{Young Stellar Objects (YSOs)}[YSOs]
\newabbrev\YSO{Young Stellar Object (YSO)}[YSO]
\newabbrev\PDF{Probability Distribution Function}[PDF]
\newabbrev\PSF{Point Spread Function}[PSF]
\newabbrev\LMC{Large Magellanic Cloud}[LMC]
\newabbrev\MHD{magnetohydrodynamics}[MHD]

\makeatletter
\newcommand*\bigcdot{\mathpalette\bigcdot@{.5}}
\newcommand*\bigcdot@[2]{\mathbin{\vcenter{\hbox{\scalebox{#2}{$\m@th#1\bullet$}}}}}
\makeatother

\makeatletter
\renewcommand\maybe@space@{%
	\maybe@ictrue % <= this is new
	\expandafter   \@tfor
	\expandafter \reserved@a
	\expandafter :%
	\expandafter =%
	\nospacelist
	\do \t@st@ic
	\ifmaybe@ic % <= this is new
	\space
	\fi
}
\makeatother
\title[Wind Bubbles in HII Regions]{The Geometry and Dynamical Role of Stellar Wind Bubbles in Photoionised HII Regions}

\author[Geen et al]{
Sam Geen$^{1}$\thanks{E-mail: s.t.geen@uva.nl} \orcid{0000-0002-3150-2543},
Rebekka Bieri$^{2}$ \orcid{https://orcid.org/0000-0002-4554-4488},
Joakim Rosdahl$^{3}$ \orcid{https://orcid.org/0000-0002-7534-8314},
Alex de Koter$^{1}$  \orcid{https://orcid.org/0000-0002-1198-3167}
\\
$^{1}$ Anton Pannekoek Institute for Astronomy, Universiteit van Amsterdam, Science Park 904, 1098 XH Amsterdam, The Netherlands\\
$^{2}$ Max-Planck-Institute for Astrophysics, Karl-Schwartzschild-Strasse 1, Garching, Germany\\
$^{3}$ CRAL, Universit\'e de Lyon 1, CNRS UMR 5574, ENS-Lyon, 9 Avenue Charles Andr\'e, 69561, Saint-Genis-Laval, France\\
}

\date{\today}

\begin{document}
\label{firstpage}
\pagerange{\pageref{firstpage}--\pageref{lastpage}}
\maketitle

\begin{abstract}
Winds from young massive stars contribute a large amount of energy to their host molecular clouds. This has consequences for the dynamics and observable structure of star-forming clouds. In this paper, we present radiative magnetohydrodynamic simulations of turbulent molecular clouds that form individual stars of 30, 60 and 120 solar masses emitting winds and ultraviolet radiation following realistic stellar evolution tracks. We find that winds contribute to the total radial momentum carried by the expanding nebula around the star at 10\% of the level of photoionisation feedback, and have only a small effect on the radial expansion of the nebula. Radiation pressure is largely negligible in the systems studied here. The 3D geometry and evolution of wind bubbles is highly aspherical and chaotic, characterised by fast-moving ``chimneys'' and thermally-driven ``plumes''. These plumes can sometimes become disconnected from the stellar source due to dense gas flows in the cloud. Our results compare favourably with the findings of relevant simulations, analytic models and observations in the literature while demonstrating the need for full 3D simulations including stellar winds. However, more targeted simulations are needed to better understand results from observational studies.
\end{abstract}


%%%%%%%%%%%%%%%%%%%%%%%%%%%%%%%%%%%%%%%%%%%%%%%%%%

%%%%%%%%%%%%%%%%% BODY OF PAPER %%%%%%%%%%%%%%%%%%

\section{Introduction}
\label{introduction}

Stars form from gas that collapses under gravity. The more massive stars eject large quantities of energy and mass over their lifetimes. This process is often called ``feedback'', because it has the capability to regulate future star formation by driving away or evaporating dense gas. These feedback processes include radiation at multiple wavelengths, jets and other protostellar outflows, winds and supernovae. Feedback is a ``multi-scale'' process, i.e. it affects a large range of spatial scales. On the smallest scales, stars regulate their own formation. This has been modelled by \cite{Kuiper2018} for massive stars above 50 \Msolar, and by \cite{Bate2019} for clusters of less massive stars (up to 10 \Msolar), amongst others. Stars also regulate gas flows on cloud scales of 1-100 pc \citep[see review by][and later references in this Section]{Dale2015a}, collectively cause the thermal phases of the interstellar medium \cite[e.g.][]{McKee1977,Gatto2017} and even drive flows out of galaxies \citep[see review on galactic winds by][]{Veilleux2005} and shape the ionisation state of the Universe \citep[e.g.][]{Rosdahl2018}. 

In this paper we focus on the interaction between two categories of feedback processes on cloud scales, namely high-energy radiation and winds, in the first Myr of the main sequence of massive stars. The first, photoionisation feedback, is driven by the ionisation of interstellar material by photons above the ionisation energy of hydrogen (13.6 eV). This heats the gas to approximately $10^4$ K, which creates a pressure difference between the ionised gas and neutral material outside, causing the ionised bubble to expand \citep{Oort1955}. Photons can also affect the gas via direct radiation pressure when it interacts with gas and dust in the ambient medium \citep{Mathews1967}. 

The second, stellar wind feedback, is the ejection of material from the surface of the star through radiation pressure exerted on the gas in the star's atmosphere. %Around massive stars, this material can be accelerated to thousands of km/s, according to models by, e.g., \cite{Vink2011}. This has been confirmed observationally, for example by \cite{Doran2013} in the VLT-FLAMES Tarantula Survey. However, it is not immediately obvious how these processes interact, and which process has the biggest impact on its surroundings.
Around young massive stars, this material can be accelerated to thousands of km/s, according to models of, e.g., \citet{Castor1975,1985ApJ...288..679A,2004A&A...417.1003K,2012A&A...537A..37M}. This has been confirmed observationally by e.g. \citet{1989A&AS...79..359G,1990ApJ...361..607P,2016MNRAS.458..624C}. However, it is not immediately obvious how radiation and wind feedback interact, and which process has the biggest impact on its surroundings. In this paper, we analyse 3D feedback on scales of 1-100\,pc from single massive stars of up to 120 \Msolar in the first million years of their life using magnetohydrodynamic (MHD) simulations, disentangling and quantifying the effects of photoionisation, radiation pressure and stellar winds. The interplay of these modes of feedback, together with initial conditions, create complex 3D geometries of cold neutral, hot ionized, and ultra-hot x-ray emitting gas characterized by kinetic energy-driven ‘chimneys’ and thermally-driven ‘plumes’ that may be disconnected from the stellar source. The wind bubble and photoionised gas combined are referred to as an \HII region, since both contain ionised hydrogen. We will use this definition throughout the paper.

\subsection{Analytic Models for Winds and Photoionisation}

Early analytic work by \cite{KahnF.D.1954}, \cite{SpitzerLyman1978}, \cite{Whitworth1979} and others confirms that photoionisation feedback is capable of driving gas flows into the interstellar medium. The same is true for analytic work focussed on adiabatic stellar wind bubbles (bubbles of hot gas driven by stellar winds), by \cite{Avedisova1972}, \cite{Castor1975} and \cite{Weaver1977}.

More recently, authors have studied the interaction between these two processes in more detail. Analytic calculations by \cite{Capriotti2001} suggest that winds are not likely to have a significant dynamical effect on the expansion of \HII regions compared to photoionisation. \cite{Krumholz2009} argue that leakage from stellar wind bubbles due to fragmentation or inhomogeneities in the shell further reduces the dynamical input from winds. \cite{Haid2018} confirm this with analytic models and simulations using a uniform dense, neutral background, but argue that once the wind has entered the already ionised interstellar medium, photoionisation cannot drive further outflows and winds are required. This is because the warm ionised medium outside the cloud has a similar ionisation state and temperature to the \HII region, and so there is no pressure difference across the \HII region radius. 

\cite{Geen2019} use analytic models of winds and radiation expanding into a power law density field, as is expected for recently-formed massive stars \citep{Lee2018a}. In this model, winds only become more important than photoionisation close to massive stars ($<~$0.1 pc). This is because the energy or momentum from winds is spread across a smaller surface area.

\cite{Rahner2017} argue that winds from massive stellar clusters do contribute a significant fraction of the force acting on the \HII region, peaking at around 3 Myr when massive stars begin to emit stronger winds during the Wolf-Rayet phase. However, \cite{Silich2017} argue that this depends on the ability for the wind bubbles around individual stars to merge, with isolated wind bubbles being less effective. \cite{Fierlinger2016} argue that winds deposit around 2-3 times the energy from supernovae into the surrounding material, and further that careful modelling of the mixing of hot and cold gas at the bubble interface is crucial for determining how much of the energy deposited by winds is lost to radiative cooling.

\subsection{Structure and Observability}

\cite{Harper-Clark2009}, \cite{Yeh2012} and \cite{Yeh2013} construct quasi-static 1D models of \HII regions including photoionisation, radiation pressure and winds. They argue that nearby observed \HII regions are consistent with models in which winds are not dynamically significant. \cite{Pellegrini2007}, \cite{Pellegrini2011} and \cite{Pellegrini2012} argue that winds are required to explain the observed structure of these regions. This is dynamically significant since winds shape the density of the photoionised region, which in turn affects the thermal pressure of the photoionised gas. They also argue that there are certain regions, such as the Orion Veil nebula, for which pressure equilibrium has not been reached and which are not well described by such quasi-static models. 

\cite{Guedel2007} find extended X-ray emission inside the Orion Veil nebula, arguing that winds fill the \HII region. \cite{Pabst2019} find that the shell around the Orion Veil nebula travels faster than the sound speed in ionised gas, while the cooling from X-ray emission is low, and thus the expansion of the region is best explained by adiabatic wind models versus a photoionisation-driven model. \cite{Kruijssen2019} also argue that the dispersal of molecular clouds by adiabatic stellar winds and photoionisation should happen at the same rate. 

\subsection{Numerical Simulations}

\rev{These (semi-)analytic analyses do not always agree, since by their nature they rely on simplified assumptions, both to reduce computational cost and to create solutions that are easy to follow from first principles. Their geometry is often simplified to enable this, neglecting the complex 3D distribution of structures in star-forming regions in galaxies or using only simple clumping factors to account for their omission. There is value in this genre of models for their ability to populate a large parameter space quickly, but they also have innate limitations that must be overcome in order to create more realistic theoretical models for feedback in star-forming regions.} 
	
\rev{There is thus a need for more comprehensive, if costly, self-consistent \rev{3D} radiation-hydrodynamic simulations to study this phenomenon. When stellar winds are included, the cost of these simulations is increased dramatically. Due to the high temperature of wind bubbles ($10^6$ to $10^8$ K, or even higher), satisfying the Courant condition forces the timestep of the hydrodynamic simulations to be much lower than for simulations with just photoionisation (with characteristic temperatures of $\sim10^4$ K). As smaller spatial scales are resolved, this timestep drops further. The problem of combining stellar winds and photoionisation has thus to this point still not been explored fully.}

In this paper we focus on molecular cloud scales. On protostellar scales, protostellar jets and outflows are the main feedback sources, with winds and ionising radiation expanding into the cavities created by these processes \citep{Kuiper2018}, while at larger scales \citep{Agertz2013,Gatto2017} winds add to the thermal pressure in hot gas in the galaxy and reduce the Galactic \SFE. \rev{\cite{Grudic2020} include winds and radiation in simulations of very massive ($> 10^6~$\Msolar) clouds, considering star particles to be well-sampled populations for the purpose of feedback.} Our work should thus be seen as a bridge between these scales, tracing the flow of wind-driven structures from sub-parsec to $\sim100~$pc scales.

Simulations of wind outflows on cloud scales by \cite{Rogers2013} and \cite{Rey-Raposo2017} demonstrate that winds escape preferentially through low-density channels, reducing their effectiveness at dispersing clouds. \cite{Dale2014}, who for the first time include both photoionisation and wind feedback with self-consistent star formation on a molecular cloud scale, find that the dynamical role of winds is small compared to photoionisation. \cite{Mackey2013} and \cite{Mackey2015} argue, using simulations of stars moving at varying speeds with respect to the background, that winds lose most of their energy to evaporation and mixing, with photoionisation being the principal driver of \HII regions around the star. Nonetheless, emission from the interface between the wind bubble and the gas around it is an important observational tracer \citep{Green2019}.

Magnetic fields have also often been omitted from simulations with radiative and wind feedback on a cloud scale due to the additional computational cost. However, as we showed in \cite{Geen2015b}, magnetic fields are important for the structure of \HII regions since they limit the breakup of filaments and shells \citep[see also][]{Hennebelle2013}. Recent work by \cite{Wall2019} simulates self-consistent photoionisaton and winds with MHD, although since the paper focusses on resolving stellar multiplicity, in their highest resolution model they do not form stars larger than 10 \Msolar, which have weaker winds and ionising photon emission rates than more massive stars.

\subsection{Outline of this Work}

In this work, we present radiative magnetohydrodynamic simulations of photoionisation, radiation pressure and wind feedback in turbulent molecular clouds. We follow the formation of massive stars self-consistently through sink particle accretion. However, in order to isolate the effects of stellar winds in controlled conditions, we allow only one massive star to form of a pre-selected mass of either 30, 60 and 120 \Msolar. The advantage of this approach is that the source of early stellar feedback is in a realistic position within the cloud, i.e. on a gas density peak. From this star we track feedback according to a full single-star evolution model (see Section \ref{methods}). Sink particle accretion, representing the formation of lower mass stars, continues. We then follow the evolution of the wind bubble and photoionised region. Our goal is to study the complex geometry of the wind bubble, the feedback efficiencies and interactions of stellar winds and radiation.

In Section \ref{methods} we discuss the methods used to set up and run our simulations. In Section \ref{results}, we present the results of these simulations, focussing on the evolution of the wind bubble. In Section \ref{discussion}, we compare our results to analytic models and observations, and discuss the results of our simulations in the context of the wider literature. Finally, in Section \ref{conclusions} we summarise our conclusions.

\section{Numerical Simulations}
\label{methods}

% Simulations
\begin{table*}
	\centering
	\caption{List of simulations included in this paper. Cloud refers to the cloud setup used (see Table \ref{methods:cloudtable}). $M_*$ refers to the mass of the star used as a source of winds and radiation. NOFB indicates that no feedback is included. UV indicates that UV photoionisation is included. WIND indicates that stellar winds are included. PRESS indicates that radiation pressure is included. See Section \ref{methods} for a discussion of how these effects are implemented.}
	\label{methods:simtable}
	\begin{tabular}{lccccccc} % four columns, alignment for each
		\hline
		Simulation name     & Cloud             & $M_*$ / \Msolar & UV     & WIND   & PRESS  \\
		\hline
		\hline
		NOFB                & \textsc{diffuse}  & -                   & \cross & \cross & \cross \\
		\hline
		UV\_30              & \textsc{diffuse}  & 30                  & \tick  & \cross & \cross \\
		UVWIND\_30          & \textsc{diffuse}  & 30                  & \tick  & \tick  & \cross \\
		UVWINDPRESS\_30     & \textsc{diffuse}  & 30                  & \tick  & \tick  & \tick  \\
		\hline
		UV\_60              & \textsc{diffuse}  & 60                  & \tick  & \cross & \cross \\
		UVWIND\_60          & \textsc{diffuse}  & 60                  & \tick  & \tick  & \cross \\
		UVWINDPRESS\_60     & \textsc{diffuse}  & 60                  & \tick  & \tick  & \tick  \\
		\hline
		UV\_120             & \textsc{diffuse}  & 120                 & \tick  & \cross & \cross \\
		UVWIND\_120         & \textsc{diffuse}  & 120                 & \tick  & \tick  & \cross \\
		UVWINDPRESS\_120    & \textsc{diffuse}  & 120                 & \tick  & \tick  & \tick  \\
		\hline
		NOFB\_DENSE         & \textsc{dense}    & -                   & \cross & \cross & \cross \\
		\hline
		UV\_120\_DENSE      & \textsc{dense}    & 120                 & \tick  & \cross & \cross \\
		UVWIND\_120\_DENSE  & \textsc{dense}    & 120                 & \tick  & \tick  & \cross \\
		UVWINDPRESS\_120\_DENSE  & \textsc{dense}    & 120            & \tick  & \tick  & \tick  \\
		\hline
	\end{tabular}
\end{table*}

% Clouds
\begin{table*}
	\centering
	\caption{List of cloud setups included in this paper, as described in Section \ref{methods}. $M_c$ is the initial cloud mass in \Msolar. $t_{ff}$ is the initial free-fall time of the cloud as a whole. $t_{sound}$ is the sound crossing time. $t_{A}$ is the Alv\'en crossing time. $t_{RMS}$ is the crossing time of the root mean square velocity of the initial turbulence of the cloud. $L_{box}$ is the box length. $\Delta x$ is the minimum cell size. $r_c$ is the characteristic radius of the central isothermal part of the cloud at $t=0$ (see Equation \protect\ref{isothermal}). $n_0$ is the central hydrogen number density. $B_{max,ini}$ is the maximum magnetic field strength in the initial seed magnetic field.}
	\label{methods:cloudtable}
	\begin{tabular}{ccccccccccc} % column alignment
		\hline
		Cloud name & log($M_c$ / M$_\odot$) & $t_{ff}$ / Myr & $t_{ff}/t_{sound}$ & $t_{ff}/t_{A}$ & $t_{ff}/t_{RMS}$ & $L_{box}$ / pc & $\Delta x_{min}$ / pc & $r_c$ / pc & $n_0$ / cm$^{-3}$ & $B_{max,ini}$ / \textmu$\mathrm{G}$\\
		\hline
		\textsc{diffuse} & 4                & 4.22            & 0.15               & 0.2            & 2.0              & 122            & 0.03 & 2.533  & 823.4 & 3.76 \\
		\textsc{dense}   & 4                & 0.527           & 0.075              & 0.2            & 2.0              & 30.4           & 0.03 & 0.6335 & 52700 & 60.1 \\
	\end{tabular}
\end{table*}

In this Section we describe the numerical setup of the simulations used in this paper (see Table \ref{methods:simtable} for a full list). Each of the simulations describes an isolated molecular cloud with an initial turbulent velocity field, magnetic field, self-gravity and stellar feedback. All of the simulations are performed with the radiative magnetohydrodynamic Eulerian \AMR code \textsc{RAMSES} \citep{Teyssier2002,Fromang2006,Rosdahl2013}. The total CPU time used in these simulations was approximately 500 khr. More specific details about the setup of the code can be found in the Data Availability statement in Section \ref{data-management}.

\subsection{Initial Conditions and Refinement Criteria}

We use two sets of initial conditions, one of a diffuse cloud similar to the nearby Gould Belt clouds, and one denser \citep{Geen2017}. In both of them, we define a cloud with an initially spherically symmetric density profile $n(r)$ defined by
\begin{equation}
n(r) = n_0 / (1 + (r/r_c)^2)
\label{isothermal}
\end{equation}
where $n_0$ and $r_c$ are the central hydrogen number density and characteristic radius, respectively. This profile is imposed out to a radius $r_{ini}=3 r_c$, where $n(r_{ini})=0.1 n_0$. Outside this, a uniform sphere is imposed up to $2~r_{ini}$, with a density 0.1 times that just inside $r_{ini}$, or 0.01 $n_0$, to provide a reservoir of material to accrete onto the cloud. The initial temperature inside $2~r_{ini}$ is set to 10 K. Outside this radius, the hydrogen number density is set to 1 cm$^{-3}$ and the temperature to 8000 K. The total length of the cubic volume simulated $L_{box} = 16~r_{ini}$. Note that the cloud evolves significantly between the start of the simulation and the time the first star forms.

There are two clouds used in this study, both with an initial mass of $10^4$ \Msolar. One is a cloud similar to the nearby Gould belt as established in \cite{Geen2017} by comparing the column density distributions from our simulated clouds and the observed clouds. The other is denser, to test the effects of feedback in different environments. These are, respectively, the ``L'' and ``S'' clouds in \cite{Geen2017}. We list the properties of both of these clouds in Table \ref{methods:cloudtable}.

In the initial conditions, we impose a supersonic turbulent velocity field over the cloud. We do not apply further turbulent forcing to the cloud. Each cloud has a global free-fall time $t_{ff} \equiv \sqrt{3 \pi / 32 G \rho_{av}}$, defined by the average density of the isothermal sphere $\rho_{av}$ inside $r_{ini}$. The radius of the cloud is set via the sound crossing time $t_{sound}$ for a fiducial neutral gas sound speed of 0.28 km/s. The balance of turbulence and gravity is set via the turbulent root mean square (RMS) velocity $V_{RMS}$, which has a crossing time $t_{RMS}$. The magnetic field strength is set via the Alv\'en wave crossing time $t_{A}$.

The magnetic field is initially oriented along the $x$ direction. We calculate a maximum initial magnetic field strength $B_{max,ini}$ using the value of the Alv\'en wave crossing time $t_{A} \equiv r_c \sqrt{\rho_0} / B_{max,ini}$, where $\rho_0 = n_0 m_H / X$ for a hydrogen mass of $m_H$ and hydrogen mass fraction $X$. We assign $B_{max,ini}$ to the density peak at the centre of the cloud. We calculate the gas column density $\Sigma_x$ along each line of sight in the x direction, and calculate the magnetic field strength $B_x$ of each cell along a given line of sight as
\begin{equation}
B_x = B_{max,ini}(\Sigma_x / \Sigma_{max,ini})
\end{equation}
where $\Sigma_{max,ini}$ is the initial maximum column density in the $x$ direction, which intersects the density peak of the cloud. At $t>0$, the magnetic field then evolves with time according to the HLLD scheme described in \cite{Fromang2006}. The maximum magnetic field strength grows considerably larger than $B_{max,ini}$ over time.

We ``relax'' the clouds by running the simulations without self-gravity for 0.5 $t_{ff}$, in order to mix the turbulent velocity and density fields, since the density field is initially spherically symmetric \citep[see][amongst others]{Klessen2000,Lee2016a}. After $0.5 t_{ff}$ we apply self-gravity to the cloud, which allows the gas to collapse to form sink particles as described in Section 
\ref{methods:sinks}.

We trace the gas dynamics on an octree mesh that refines adaptively when certain conditions are met. Every time a cell at level $l$ fulfils certain criteria, it subdivides itself into 8 child cells at level $l+1$. The cell size is given by $\Delta x = L_{box} / 2^l$.

We minimally refine everywhere up to level 7, giving a cube with $2^7=128$ cells on a side. Everywhere inside a sphere of diameter $8 r_{ini}$ we fully refine up to level 9, i.e. two further levels. Finally, any gas cell that is ten times denser than the Jeans density\footnote{Jeans density $\rho_J \equiv (c_s/\Delta x)^2/G$, where $c_s$ is the sound speed in the cell and $\Delta x$ is the length of the cell. Note that this does not include support from magnetic pressure.} or has a mass above 0.25 \Msolar anywhere in the simulation volume is refined, down to a minimum cell size $\Delta x_{min}$ of 0.03 pc. This corresponds to level 12, or level 10 in the `\textsc{dense}' clouds. %$L_{box}$ and minimum cell size $\Delta x_{min}$ are listed with other cloud properties in Table \ref{methods:cloudtable}.

\subsection{Cooling and Radiative Transfer}
\label{methods:cooling}

We track the propagation of radiation across the full \AMR grid using the M1 method \citep{Rosdahl2013}. The radiation is coupled to the gas via photoionisation, dust absorption, and direct pressure from the transfer of photon momentum to the gas. In runs labelled ``UV'' (see Table \ref{methods:simtable}), we track extreme ultraviolet (EUV) photons above the ionisation energy of hydrogen, with photons binned into three groups bounded by the ionisation energies of HI, HeI and HeII. In these runs we do not automatically include radiation pressure. Runs with radiation pressure included are labelled ``PRESS''. In these runs, we include an additional far ultraviolet (FUV) group between 5.7 eV and 13.6 eV, which interacts only via radiation pressure on dust (see below).

In each grid cell, the code stores the photon density and flux for each group, and couples the photons to the gas at every timestep via the cooling function. Radiation travels at a reduced speed of light of 0.01 c, in order to reduce the cost of the radiation transport steps. This value is chosen to be similar to the maximum speed of stellar winds, and to ensure that the code can capture the speed of ionisation fronts in the simulation. We subcycle the radiation step, so that the hydrodynamic timestep is not limited by the (reduced) speed of light. More details on how this is done are given in \citet{Rosdahl2013}. \rev{In \cite{Bisbas2015} we use this value for the reduced speed of light in our code to compute a known test problem and find excellent agreement with other radiative transfer codes that use different techniques, with differences on the order of 1\%, or one grid cell}.

Each grid cell tracks the ionisation state of hydrogen and helium. Ionisation fractions change via photoionisation, recombination and collisional ionisation as calculated in the radiative transfer module described in \citet{Rosdahl2013}. Each group uses a ``grey'' approximation, i.e. all photons in the group are considered to have the same energy, energy-weighted cross section and number-weighted cross section, using representative values from a Starburst99 \citep{Leitherer2014} stellar population as in \citet{Geen2017}.

Radiation pressure is calculated in runs labelled ``PRESS'' according to \cite{Rosdahl2015}. Direct radiation pressure is applied for each photon absorption event in the gas. The local gas opacity to the radiation in all ionising photon groups is given by $\kappa_{abs}=10^3~Z/Z_{ref}~$cm$^{2}$/g, where $Z$ is the metallicity of the gas and $Z_{ref} = 0.02$. We use the reduced flux approximation described in \citep[][Appendix B]{Rosdahl2015} to ensure that the correct radiation momentum pressure is applied.  Each absorption event transfers momentum from the photons to the gas. In addition to the EUV bins interacting with atoms, all radiation bins interact with dust. We do not track the formation, destruction and advection of dust self-consistently, and instead assume a fixed coupling to the gas proportional to the gas metallicity and neutral hydrogen fraction. As the EUV photons are absorbed by dust grains, this reduces the budget of photons that can ionise hydrogen and helium.

Cooling rates are applied on a per-cell basis. The temperature of the photoionised hydrogen and helium evolves at each timestep following the recombination cooling and photoionisation heating functions described in \citet{Rosdahl2013}. To account for metals, we add a cooling rate $\Lambda_{metal}=\Lambda_{metal,ion} x + \Lambda_{metal,neutral} (1-x)$ where $x$ is the hydrogen ionisation fraction. $\Lambda_{metal,neutral}$ follows the cooling module of \citet{Audit2005} that uses fits to various coolants in an \ISM environment with a heating term from a radiation background below the ionisation energy of hydrogen. Cooling for photoionised metals, $\Lambda_{metal,ion}$, is described by a fit to \citet{Ferland2003}. Above $10^4$ K, we use a fit to \citet{Sutherland1993} where the cooling rate is higher than our simple photoionised model. The gas is set to solar metallicity, which we take to be $Z = 0.014$ as in \cite{Ekstrom2012}. We consider metal enrichment from a single star's winds to be negligible at this metallicity.

\subsection{Sinks and Star Formation}
\label{methods:sinks}

If a gas cell is above 10\% of the Jeans density at the highest refinement level, it is assigned to a ``clump'', i.e. a patch of dense gas. Clump peaks are identified using the ``watershed'' method, in which contours from high to low density are drawn, with clumps merged by identifying saddle points in the density field \citep{Bleuler2014}. If a clump is denser than the Jeans density at the highest refinement level, a sink particle is formed, and every timestep, 90\% of the mass in the clump above the Jeans density is accreted onto the sink particle \citep{Bleuler2014a}. This 90\% is a heuristic quantity to prevent zero or negative densities in accreting cells.

Once the total mass of all sink particles in the simulation exceeds 120 \Msolar, we create a \rev{``stellar object''} representing a massive star of mass $M_*$ and attach it to the most massive sink, as in \citet{Geen2018}. The rest of the mass is considered to be stars below 8 \Msolar, which produces negligible high-energy radiation or winds. \rev{None of the mass in the sink is removed, and this stellar object is simply a book-keeping tool to track the age and number of massive stars in the simulation. This age is used by the stellar evolution and feedback model to calculate the radiative and mechanical feedback from the star. This radiation, mass, momentum and energy is deposited at the position of the host sink every timestep. We use this method, rather than assigning a fixed position to the stellar object, because as \citet{Geen2019} show, the density profile around the star has a dramatic influence on the dynamics of the radiation and wind feedback.} 

All sinks continue to accrete to a larger mass if they continue to fulfil the criteria described above. In this paper we run simulations where $M_*$ is either 30, 60 or 120 \Msolar. We also run simulations with no stellar object (i.e. $M_* = 0$). We pick these masses to sample a range of masses of stars where winds are expected to have a greater effect \citep{Geen2019}. \rev{Based on the estimates of \cite{Weidner2009}, these stellar masses are possible given the reservoir of gas in our simulation, although a 120 \Msolar star is significantly more likely to be formed in a more massive cloud}. We do not include more stellar masses in this study due to limits in the masses included in the stellar tracks we use (see Section \ref{methods:evolution}), although these stars have been reported \citep[e.g.][]{Crowther2010,Bestenlehner2010}. \rev{In each simulation we only allow one massive star to form, in order to study in detail the response of the gas to winds and radiation from a single source.}

\subsection{Stellar Evolution and Feedback}
\label{methods:evolution}

We implement feedback from the massive star as the emission of UV radiation and winds. The star is considered to start on the main sequence from the moment of formation, since we do not have the resolution to properly track the protostellar phase, which is typically $\sim 10^5$ years \citep[e.g.][]{Hosokawa2009}. The radiation and winds are emitted from the position of the sink that the star is attached to.

We follow the evolution of massive star models at solar metallicity ($Z=0.014$) computed using the Geneva model \citep{Ekstrom2012}, assuming the stars are rotating at 0.4 of the critical velocity. For completeness, we show the resulting photon emission rates and wind properties in Appendix \ref{appendix:stellartracks}. Though in the current paper we focus on the early stage of stellar evolution up to 1 Myr, we include a description of our stellar evolution tracks including older and less massive stars than the ones featured in this paper. \rev{We stop after 1 Myr due to computational cost, although by this time in all of the simulations, ionising radiation has begun to escape the simulation volume.}

At each timestep we deposit radiation and winds onto the grid. The number of photons emitted per unit time in each radiation group is calculated using individual stellar spectra extracted from \textsc{Starburst99} \citep{Leitherer2014}, using the Geneva model as inputs for the stellar atmosphere calculations. To calculate the number of photons emitted between time $t$ and $t+\Delta t$, we interpolate linearly between pre-computed tables for each photon group at intervals of 5 \Msolar.

We inject winds every timestep in the same fashion. Stellar mass loss rates $\dot{m}_w$ and escape velocities $v_{esc}$ are taken from \citet{Ekstrom2012}.
%The Geneva models calculate mass-loss rates according to the prescription of \cite{Vink2000} and \cite{Vink2001} and scale them by 0.85. 
Note that mass loss rates are uncertain by a factor of 2-3 \citep[e.g.][]{2012A&A...537A..37M,Smith2014,Puls2015,Keszthelyi2017}, and these models should be considered in this light.

We convert the escape velocity $v_{esc}$ at the stellar surface to terminal wind velocities $v_w$ using the corrections given in \citet{Gatto2017}. We list these here for clarity. We first calculate an ``effective'' escape velocity $v_{eff}$,
\begin{equation}
v_{eff}^2  =  (1 - \Gamma_e) v_{esc}^2~,
\end{equation} where $\Gamma_e$ is the Eddington factor, described in \citet{Lamers1993} as a correction factor to the Newtonian gravity set by radiation pressure on free electrons from the star. $\Gamma_e$ is given by
\begin{equation}
\Gamma_e = \frac{\sigma_e \sigma_{SB} T_e^4}{g c},
\end{equation}
where $\sigma_{SB}$ is the Stefan-Boltzmann constant, $g$ is the surface gravity, $c$ is the speed of light,
% Joki asked me to define the speed of light, it's not my fault
 and $\sigma_e$ is the cross-section for electron-photon scattering per unit mass given by
\begin{equation}
\sigma_{e}= 0.4 (1 + I_{He}Y_{He}) / (1 + 4 Y_{He})~\mathrm{cm^2 / g}
\end{equation}
where $I_{He}$ is the number of free electrons per He atom or ion, and $Y_{He}$  is the Helium abundance by number (approx 0.1). $I_{He}$ is zero below an effective surface temperature $T_e = 10^4~$K, 2 above $T_e = 2.5\times10^4~$K, and 1 otherwise. All of these parameters are calculated for each timestep of each of the different Geneva stellar tracks used in this study. $\Gamma_e$ is typically between 0.4 and 0.95 \citep{Vink2011}. 

We divide massive stars into different classifications as in \citet{Crowther2007} and \citet{Georgy2012}. Stars with $T_e > 10^4~$K and a surface hydrogen mass fraction of less than 0.3 are Wolf Rayet (WR) stars. Stars below $T_{RSG} = 5000~$K are Red Supergiants (RSG). Stars above $T_{BSG} = 8700~$K (up to $10^4~$K) are Blue Supergiants (BSG) and stars between 5000 K and 8700 K are Yellow Supergiants (YSG). Stars that do not fall into these categories are OB stars. We subdivide WR stars into categories WNL and WNE, and WC and WO, depending on the surface abundances of H, He, C, N and O. For these stars, $v_w$ is given using a clamped linear interpolation
\begin{equation}
\begin{aligned}
v_w &= v_0 \mathrm{~if~}~T_e < T_0 \\
v_w &= v_1 \mathrm{~if~}~T_e > T_1 \\
v_w &= v_0 + (v_1 - v_0) \times (T_e - T_0) / (T_1 - T_0)~\mathrm{~otherwise}
\label{kwind}
\end{aligned}
\end{equation}
where $v_0$ and $v_1$ are reference wind velocities in km/s. $T_0$ and $T_1$ are surface temperatures in units of $10^4$ K. For OB stars we use ($v_0,~v_1,~T_0,~T_1$) = ($1.3~v_{eff},~2.45~v_{eff},~1.8,~2.3$) . For WO and WC stars we use ($v_0,~v_1,~T_0,~T_1$) = ($700,~2800,~2.0,~8.0$), and for WNL and WNE stars ($v_0,~v_1,~T_0,~T_1$) = ($700,~2100,~2.0,~5.0$).

For RSG stars, we use $v_w = 10 \mathrm{~km/s~} \times (L/L_{ref})^{1/4}$, where $L_{ref} \equiv 3 \times 10^4 L_{\odot}$. For YSG stars, we use  
\begin{equation}
\begin{aligned}
v_w = 10 \mathrm{~km/s~} \times 10^{[(\mathrm{log}(T_e) - \mathrm{log}(T_{RSG})) / (\mathrm{log}(T_{BSG}) - \mathrm{log}(T_{RSG}))]},
\end{aligned}
\end{equation}
in order to fit the argument of \citet{Gatto2017} that the geometric mean velocity in this range is 50 km/s, while RSG winds are typically somewhere above 10 km/s and BSG winds are 100 km/s. Mass loss rates from RSG and YSG stars are comparable to other massive stars, but the $v_w$ is lower. Thus the momentum and energy deposition rates from stellar winds are typically much weaker for these stars than OB or WR stars. We further note that this field is subject to ongoing study, and as such this model does not represent the final word in winds from massive stars.

We force the cells around the sink particle with the massive star to be at the highest refinement level. We inject winds into a 5 cell radius around the sink. Mass, momentum and energy are injected evenly in all cells inside this radius. The injected momentum and energy are calculated as $\dot{m}_w v_w~\mathrm{d}t$ and $\frac{1}{2} \dot{m}_w v_w^2~\mathrm{d}t$ respectively, where $\mathrm{d}t$ is the timestep. When injected into a cell with a low density, the injected mass and momentum dominate, and the solution becomes free-streaming. If the cell has a high density, the injected mass and momentum have less of an impact on the final velocity of the cell, and the injected energy effectively becomes thermalised. This means that the free streaming phase of the wind described in \cite{Weaver1977} appears as the wind bubble grows and the density of gas around the star is largely from the wind itself and not the pre-existing circumstellar medium.

\section{Results}
\label{results}

In this Section we present the results of the simulations and explore the effect that winds have on the photoionisation region. We begin with an overview of the simulations. We then discuss the influence that winds have on bulk properties of the system. Finally, we look in detail at the evolution of the wind bubble itself, and on the role that radiative cooling plays in the wind bubble's evolution.

\subsection{Overview}
\label{results:global}

\begin{figure*}
	\includegraphics[width=0.5\columnwidth]{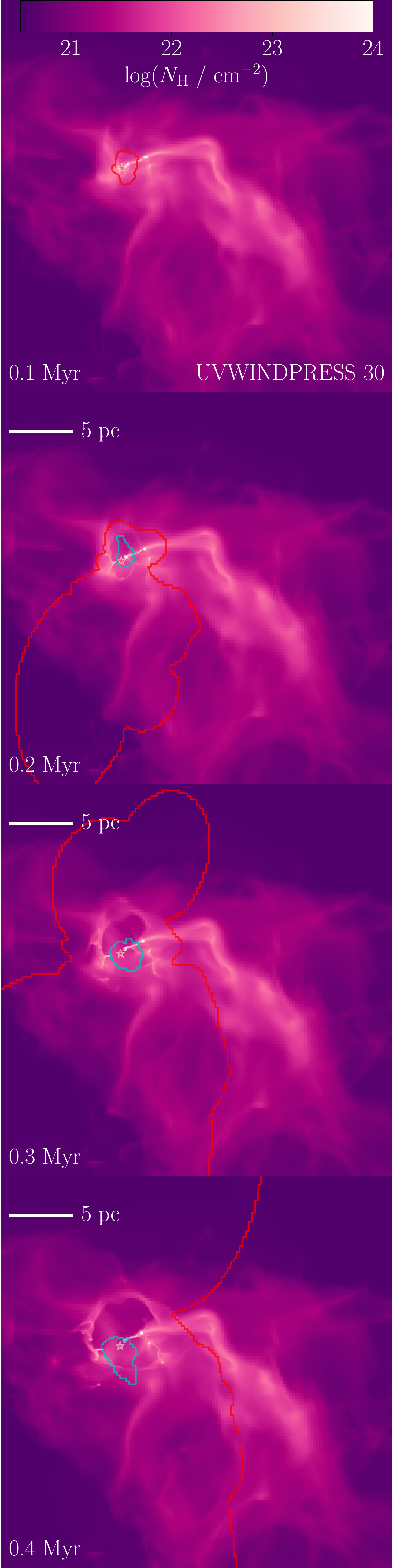}\includegraphics[width=0.5\columnwidth]{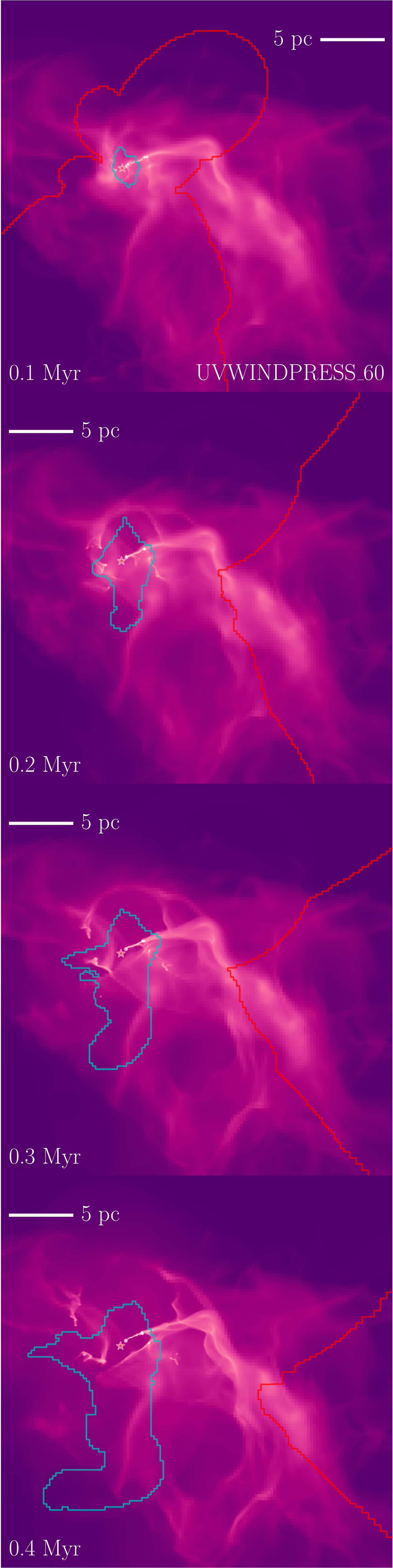}\includegraphics[width=0.5\columnwidth]{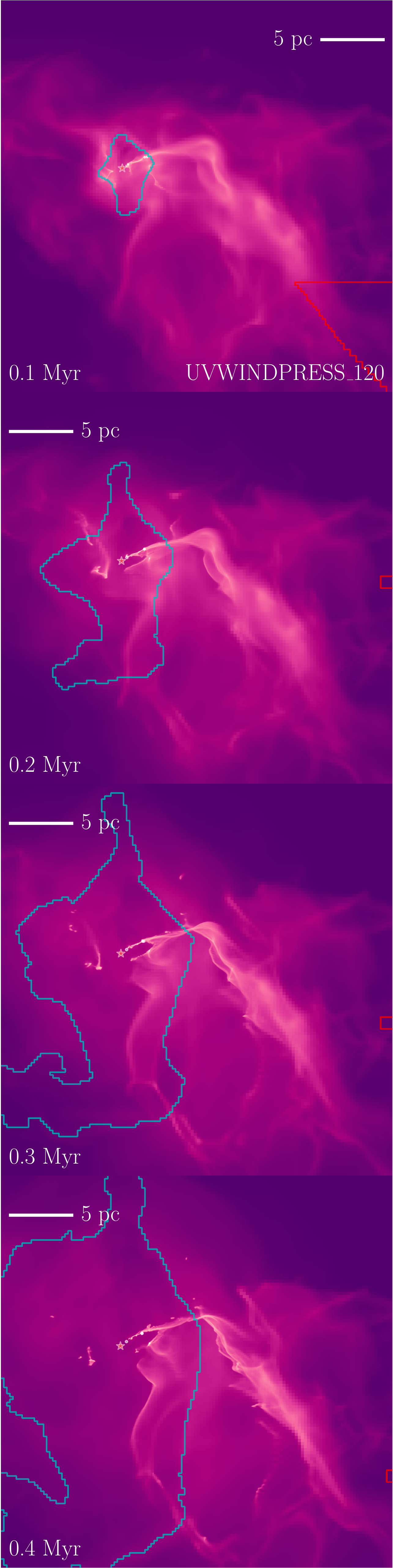}\includegraphics[width=0.5\columnwidth]{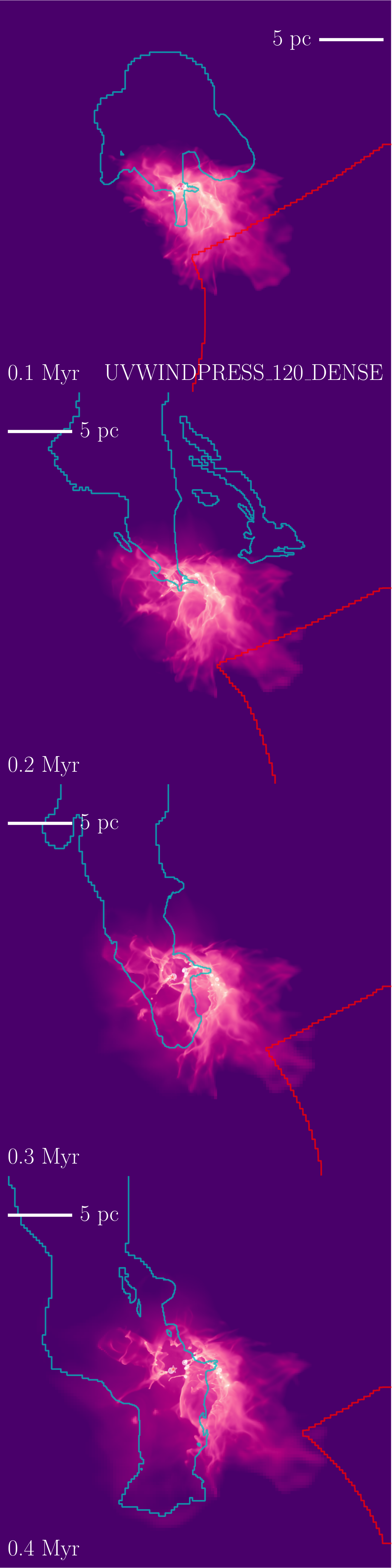}
	\caption{Maps of column density projected in the y-axis in each simulation including photoionisation, stellar winds and radiation pressure. From top to bottom are snapshots at 0.1, 0.2, 0.3 and 0.4 Myr after the star is formed. From left to right we plot results for the 30 \Msolar, 60 \Msolar star, and 120 \Msolar star, and the 120 \Msolar star in the \textsc{dense} cloud. The red contours show the extent of the ionised gas around the star. The cyan contours show the extent of the wind bubble. See Section \protect\ref{results:global} for a description of how these are defined. The ionisation front expands rapidly outwards in a champagne flow. The wind bubble expands inside this region, and has a highly aspherical geometry.}
	\label{fig:allwindsims}
\end{figure*}

%\begin{figure*}
%	\includegraphics[width=0.5\columnwidth]{../plots/vis/multiray/multirayTime_coolemission_ionemission_xrayemission2__windset_30Msun0p4Myr_zoom0p25__ywindpressonly_sequence.pdf}\includegraphics[width=0.5\columnwidth]{../plots/vis/multiray/multirayTime_coolemission_ionemission_xrayemission2__windset_60Msun0p4Myr_zoom0p25__ywindpressonly_sequence.pdf}\includegraphics[width=0.5\columnwidth]{../plots/vis/multiray/multirayTime_coolemission_ionemission_xrayemission2__windset_120Msun0p4Myr_zoom0p25__ywindpressonly_sequence.pdf}\includegraphics[width=0.5\columnwidth]{../plots/vis/multiray/multirayTime_coolemission_ionemission_xrayemission2__windset_120Msun_dense0p4Myr_zoom1p0__ywindpressonly_sequence.pdf}
%	\caption{Maps of various radiative properties of the gas projected in the y axis in each simulation including stellar winds and radiation pressure. Green shows the projected cooling rate of cool gas (below 1000 K). Orange shows the recombination rate in ionised gas. Purple shows the projected cooling rate of hot gas (above $10^6~$K). From top to bottom are snapshots at 0.1,0.2,0.3 and 0.4 Myr after the star is formed. From left to right we plot results for the 30 \Msolar star, the 60 \Msolar star, the 120 \Msolar star, and the 120 \Msolar star in the \textsc{dense} cloud. Each radiative cooling bin is normalised for the purposes of visual comparison.}
%	\label{fig:allwindsims}
%\end{figure*}

In Figure \ref{fig:allwindsims} we plot column density maps for each simulation containing photoionisation, radiation pressure and stellar winds at 0.1 Myr intervals after the formation of the star. We overplot contours showing the position of the wind bubble and the ionisation front. The wind bubble, outlined in cyan, includes all cells above $2\times10^4~$K or travelling faster than 100 km/s. These thresholds are selected to exclude photoionised gas and include cooler parts of the wind bubble, although a larger temperature threshold is possible with no change in the results. The edge of the ionisation front is shown in red. 

The photoionised \HII region expands faster than the wind bubble in all cases, rapidly escaping the cloud in a ``champagne flow'', or a directional escape of photoionised gas from the cloud described by \cite{TenorioTagle1979}. \cite{Franco1990} argue that supersonic ionisation front expansion occurs for clouds with a power law density profile with index $w > 3/2$. The wind bubble expands against the photoionised gas, which is thermalised to around $10^4~$K. Gas pushed by the wind into the surrounding medium is already photoionised. This gas mixes with the rest of the photoionised gas rather than forming a dense neutral shell.

This picture is consistent with the 1D hydrostatic spherically symmetric model we present in \cite{Geen2019}. However, there are divergences between this simple analytic picture and the full 3D simulations in this work, which we now discuss.

The wind bubble size increases with stellar mass and time. The shape of the bubble is highly aspherical, and follows channels of low density in the gas. At early stages, the wind bubble is confined, but as it expands it develops a fast-flowing chimney structure that reaches beyond the cloud. Once it does so, it develops Rayleigh-Taylor-like plumes of hot gas that extend into the interstellar medium. The chimney expands in size as the wind bubble expands and the neutral cloud is dispersed.

This chimney-and-plume structure is especially evident in the \textsc{dense} cloud. In addition, part of the plume in the dense cloud at 0.2 Myr is cut off from the rest of the wind bubble, and cools rapidly. This is caused by dense flows in the cloud temporarily cutting off part of the wind bubble from the star. This is somewhat similar to the flickering effect seen by \cite{Peters2010} in their simulations of ultracompact HII regions. This effect also occurs in the UVWIND\_120 simulation. 

The shape of the wind bubble is not stable, but changes chaotically and appears to be influenced by turbulent motions in the rest of the cloud. The wind bubble's characteristic speed (either its sound speed or wind terminal velocity, which are similar in magnitude) is around 1000 km/s or more, so it can rapidly respond to changes in the density and pressure in the surrounding gas. The growth of the total volume of the wind bubble is more stable, however.

We do not run simulations with varying magnetic fields due to the already considerable cost of these simulations. At the point the star forms, the magnetic energy in the cloud is similar to the thermal energy, although smaller than the kinetic energy in turbulence. As the \HII region expands, the magnetic energy grows, although it lags behind the thermal and kinetic energy from feedback. We will perform a more detailed parameter study of the role of magnetic fields in future work. 

\subsection{Radial Evolution of Feedback Structures}
\label{results:radius}

\begin{figure*}
	\includegraphics[width=1.9\columnwidth]{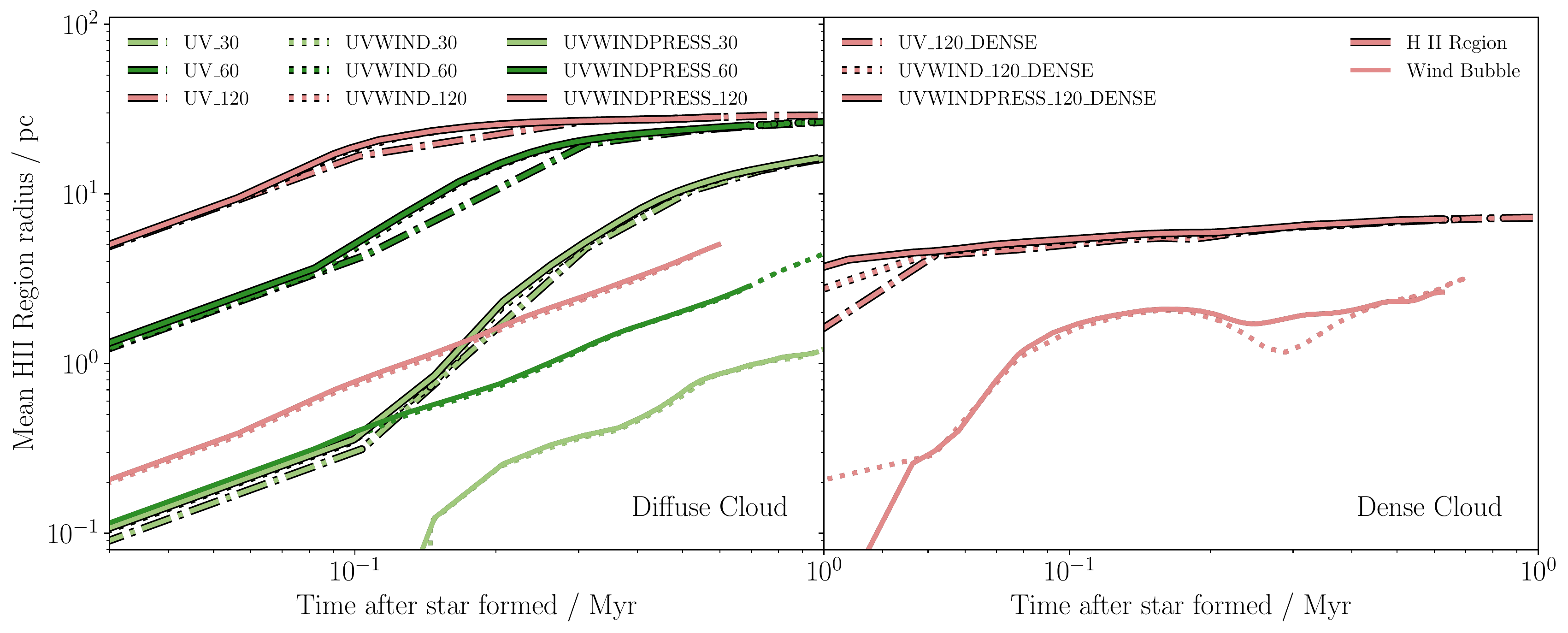}
	\caption{``Sphericised'' radius of the \HII region and wind bubble as a function of time in each simulation. Sphericised radius $r_{i,s}$ is defined via the total volume $V_i$ of the \HII region $V_i = \frac{4}{3} \pi r_{i,s}^3$. A similar radius $r_{w,s}$ is calculated for the wind bubble with volume $V_w$. $V_i$ is the total volume of all cells above a hydrogen ionisation fraction of 0.1. $V_w$ is the total volume of all cells with a bulk velocity above 100 km/s or a temperature above $2 \times 10^4$ K. Dashed-dotted lines show the \HII region radius in the UV photoionisation-only simulations, dotted lines show the result of simulations including UV photoionisation and stellar winds, and solid lines show the results of simulations including UV photoionisation, radiation pressure and stellar winds. Thick outlined lines show the radius of the ionisation front, and thin lines show the wind bubble radius in simulations including winds. On the left are the results from the \textsc{diffuse} cloud, on the right are results from the \textsc{dense} cloud. The wind bubble radius is always smaller than the ionisation front radius.}
	\label{fig:radius}
\end{figure*}

The 3D shape of the wind bubble is highly aspherical and evolves chaotically. In order to trace the bulk evolution of the wind bubble using a simple, stable diagnostic, we plot the ``sphericised'' radius of the \HII region and wind bubble in Figure \ref{fig:radius} for each simulation. The sphericised radius of the ionisation front $r_{i,s}$ is defined via the total volume $V_i$ of the \HII region or wind bubble $V_i = \frac{4}{3} \pi r_{i,s}^3$. A similar radius $r_{w,s}$ is calculated for the wind bubble with volume $V_w$. The photoionised cells and wind bubble cells counted inside $V_i$ and $V_w$ respectively are calculated as in \ref{results:global}.

The \HII regions grow slowly at first in simulations with the lower mass stars. However, in all simulations the regions eventually accelerate outwards from the star into lower density neutral cloud material, \rev{before reaching a plateau phase}. According to the expansion model of \cite{Franco1990}, this phase is unstable and the ionisation front can expand supersonically. The expansion eventually plateaus as the whole volume becomes ionised.

The radius of the \HII region in simulations with winds is only slightly larger than the radius without. The wind bubble itself is only a small fraction of the \HII region's radius in all cases, although it is a larger fraction in the \textsc{dense} cloud. We discuss the ratio between these two radii later in the paper. The flickering of the wind bubble in the \textsc{dense} cloud is seen at around 300 kyr.

Radiation pressure has almost no effect on the results. The main difference is in the volume of the wind bubble in the \textsc{dense} cloud. This is because in the UVWIND\_120_DENSE simulation, the whole wind bubble plume is cut off from the star and cools, whereas in the UVWINDPRESS\_120_DENSE simulation, only part of the plume is cut off. We attribute this to the chaotic nature of the wind bubble geometry rather than any systematic effect.

%Note that the cell size at maximum refinement is 0.03 pc, so $\sim$0.1-0.3 pc should be considered our smallest effective resolution once the injection radius of 5 cells is considered. Note also that some simulations have larger sampling intervals between outputs. The apparently slow initial evolution of the wind bubble in the 30 \Msolar simulations is due to the finite resolution of the grid, and the amount of time it takes a bubble to establish itself on the grid, even at our highest resolution.

\subsection{Radial Momentum Added to the Cloud}
\label{results:momentum}

\begin{figure*}
	\includegraphics[width=1.9\columnwidth]{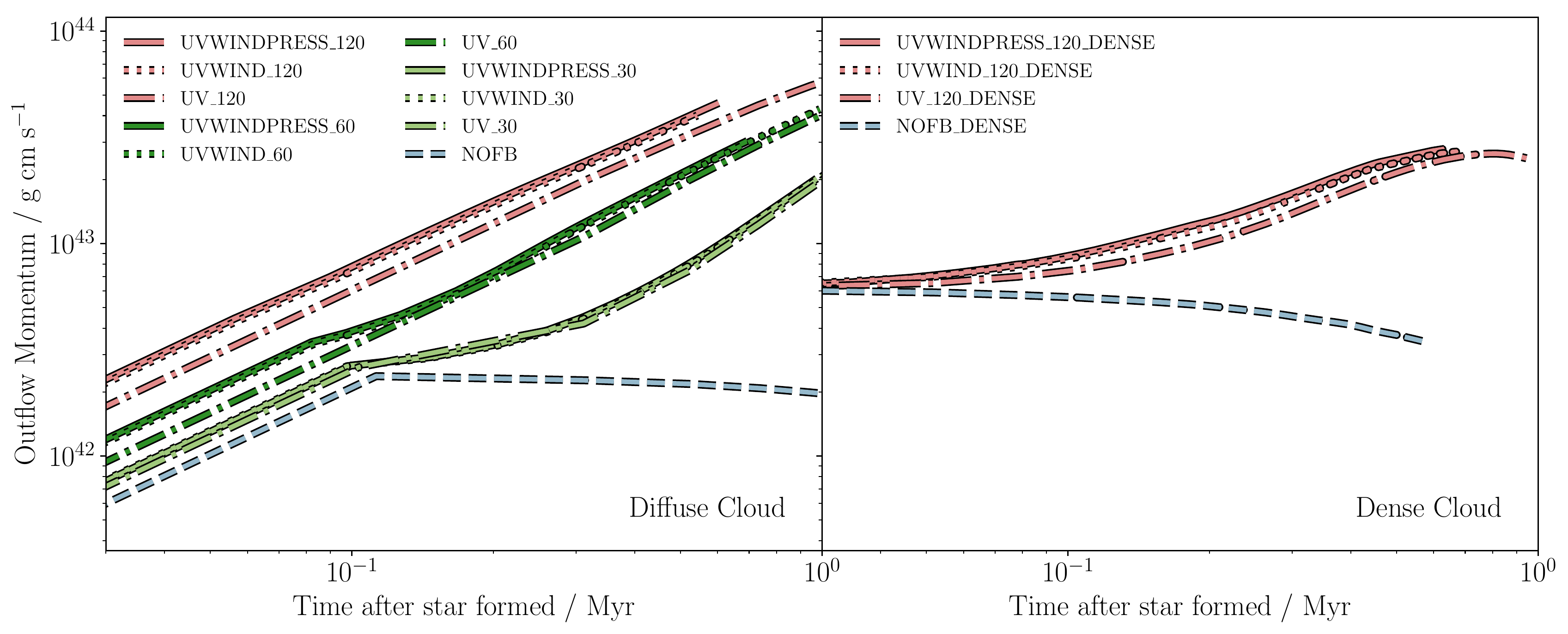}
	\caption{Radial outwards momentum from the position of the star as a function of time in each simulation. Dashed lines show the results of the NOFB simulations, dash-dotted lines show the UV photoionisation-only simulations, dotted lines show simulations with UV photoionisation and stellar winds, and solid lines show the result of simulations including UV photoionisation, radiation pressure and stellar winds. On the left are the results from the \textsc{diffuse} cloud, on the right are results from the \textsc{dense} cloud. The momentum in the NOFB run evolves due to the interplay between turbulence (some of which is oriented in a radial direction) and gravity. The contribution from winds is at most 10\% in any simulation set, while the contribution from radiation pressure is mostly negligible. The results only show momentum in directions away from the star, and ignores flows moving towards the star.}
	\label{fig:momentum}
\end{figure*}

The momentum in the cloud as a whole in flows directed away from the star in each simulation is shown in Figure \ref{fig:momentum}. This includes outflows from the star, flows driven outwards in the \HII region by feedback and the component of turbulent flows in the neutral gas in the outward radial direction. For each star, simulations containing different stellar feedback physics are shown in different line styles to demonstrate the contribution of each effect.

The boost in momentum from winds or radiation pressure in the 30 \Msolar case is negligible. The boost from winds becomes larger as the stellar mass is increased, but is no more than 10\% in the 120 \Msolar star's case. Winds never add more momentum to the flows around the star than photoionisation.

Radiation pressure provides a negligible contribution to the outflowing momentum, adding an imperceptible amount of momentum on top of the momentum added by photoionisation and winds. Our dust model is simplified, and a more complex model may give a different result. However, we do not expect a different model to change our results significantly given that radiation pressure plays a negligible role in our current simulations. Similar studies by \cite{Kim2018} and \cite{Fukushima2020} find that radiation pressure only affects star formation efficiencies in much denser regions than the ones studied here. The biggest effect would likely be to modify the budget of ionising photons for the photoionisation process as dust grains in the \HII region absorb ionising photons that would otherwise ionise hydrogen or helium (\cite{Krumholz2009} estimate a reduction of up to 27\% for a \HII region with typical Milky Way dust fractions). 

 \rev{The momentum includes all flows in the cloud in the direction away from the star, and thus contains a component of the turbulent flow, visible in the simulation without feedback (NOFB). It ignores flows moving towards or perpendicular to the star.} These flows can cause the expansion of feedback structures around stars to stall if the feedback source is weak or the cloud is dense, as we found in similar simulations in \cite{Geen2015b} and \cite{Geen2016}. This turbulent neutral gas flow momentum also accounts for the "floor" in momentum in the \textsc{dense} cloud and for the 30 \Msolar star.

\subsection{The Evolution of the Embedded Wind Bubble}
\label{results:evolutionwindbubble}

\begin{figure*}
	\centerline{\includegraphics[width=0.80\columnwidth]{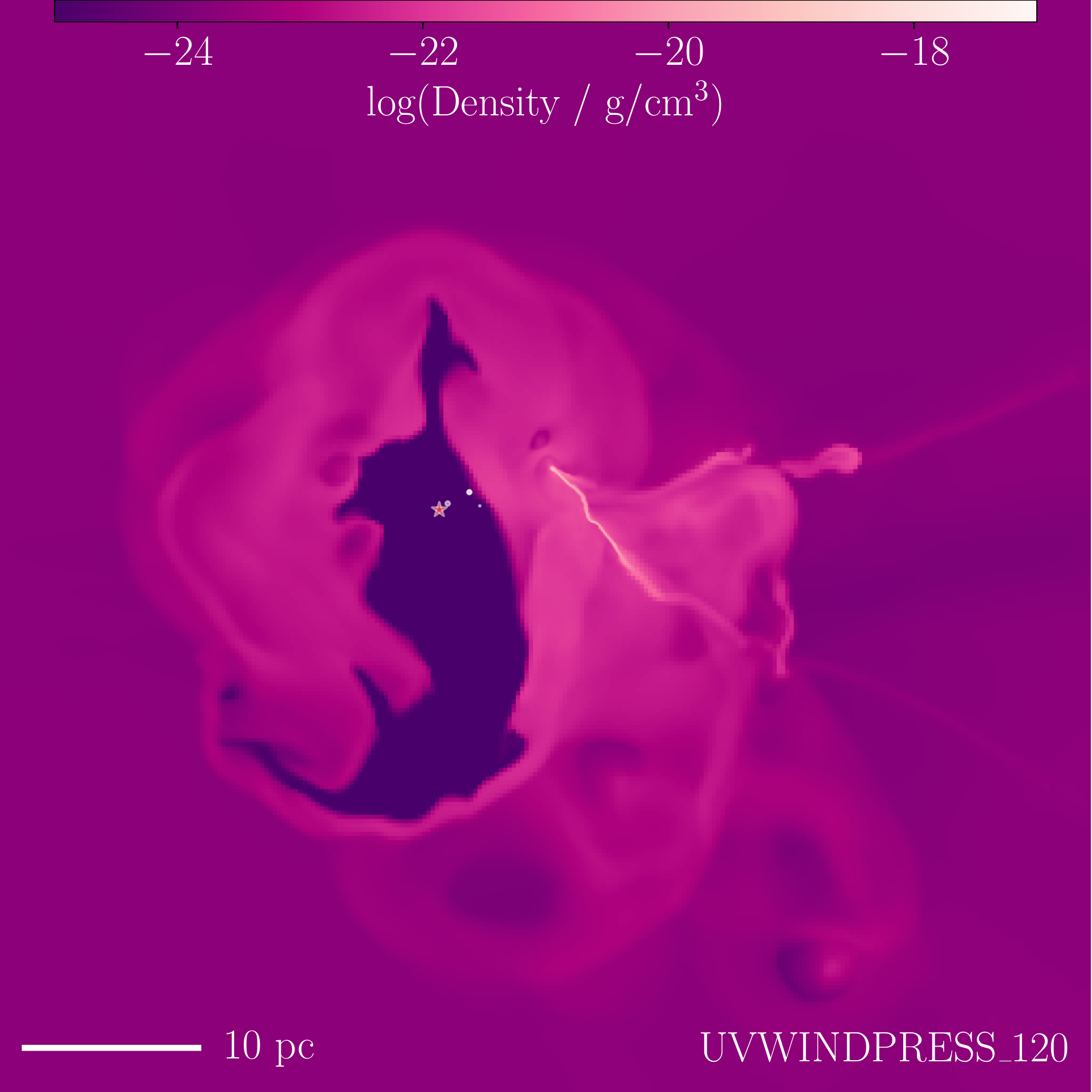} \includegraphics[width=0.80\columnwidth]{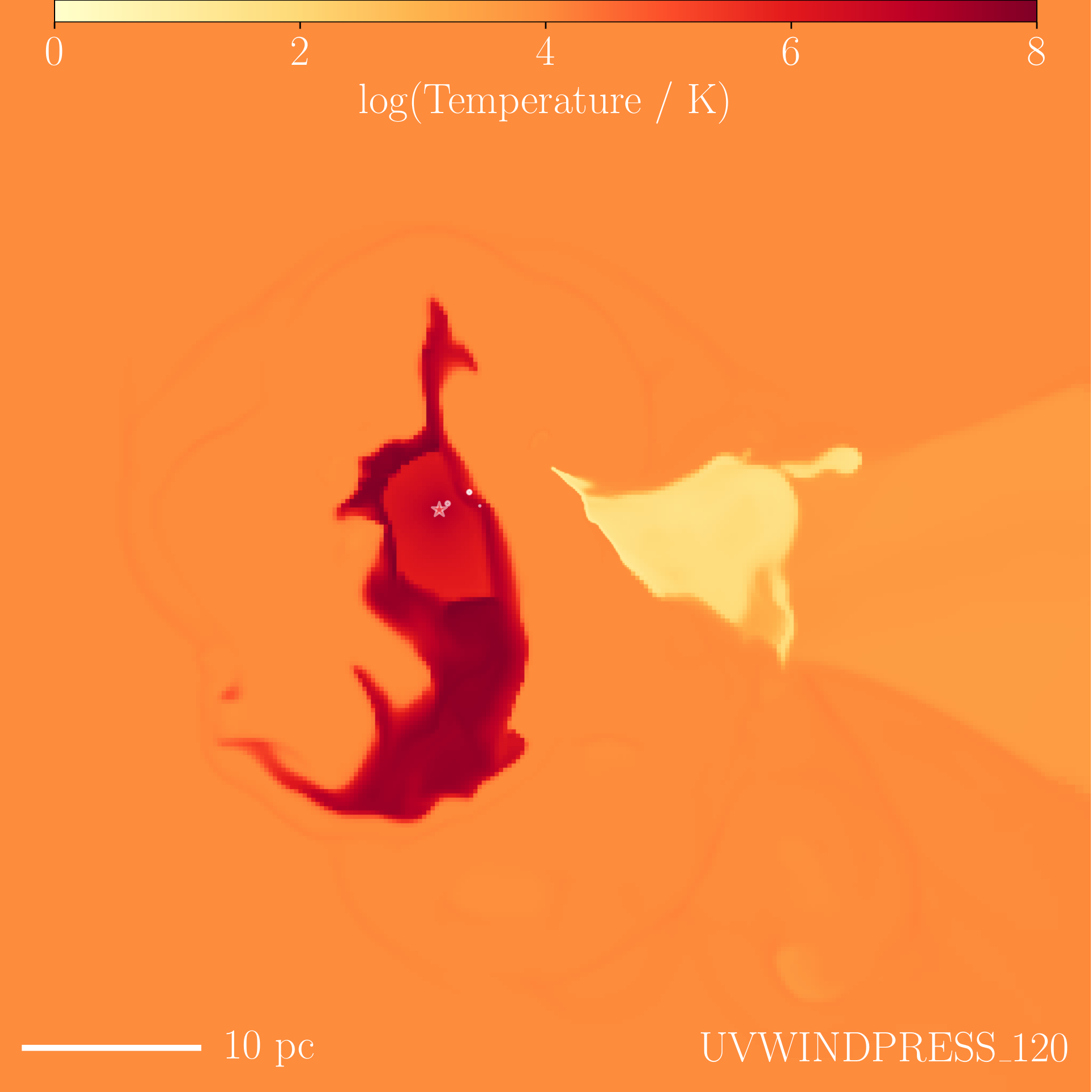}}
	\centerline{\includegraphics[width=0.80\columnwidth]{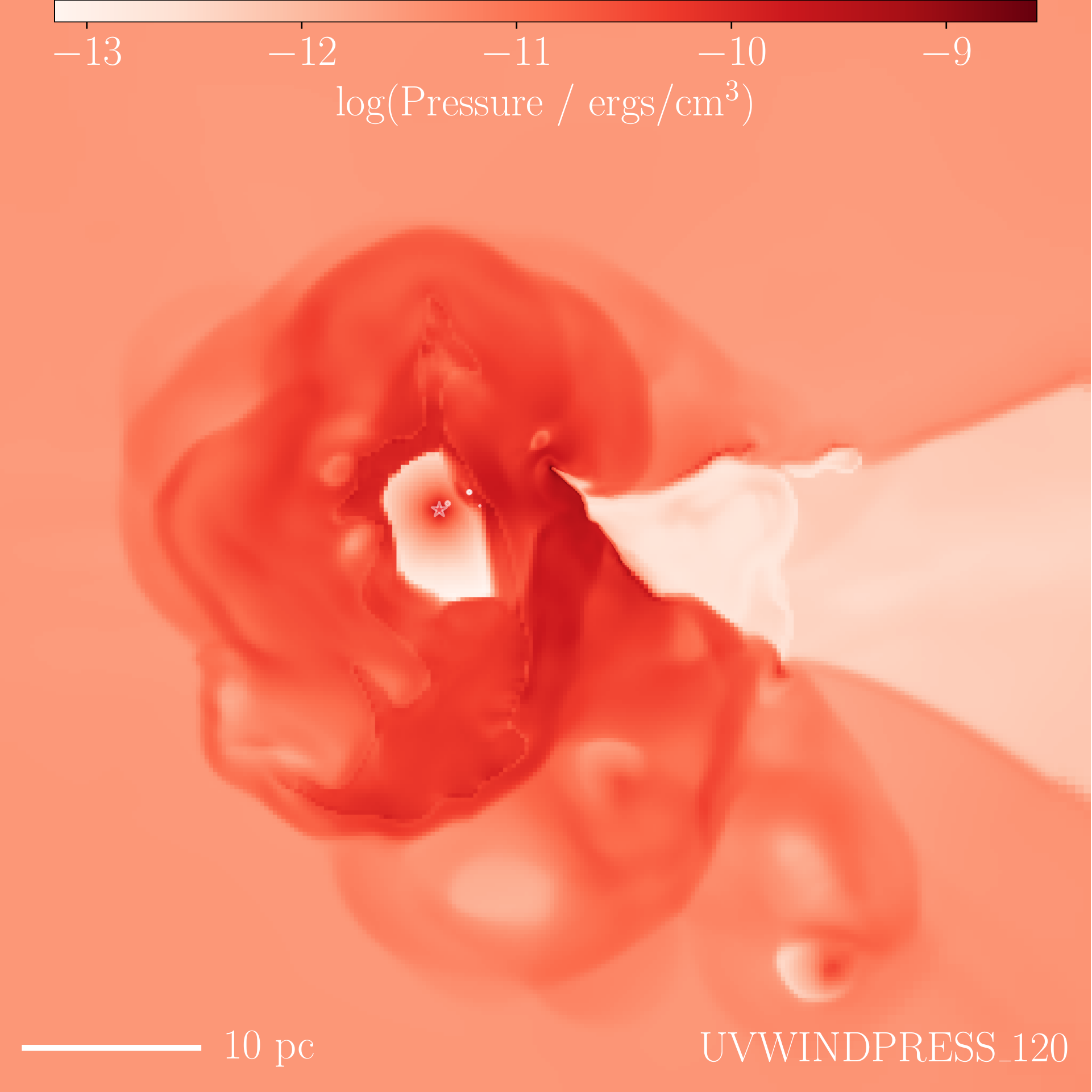}
	\includegraphics[width=0.80\columnwidth]{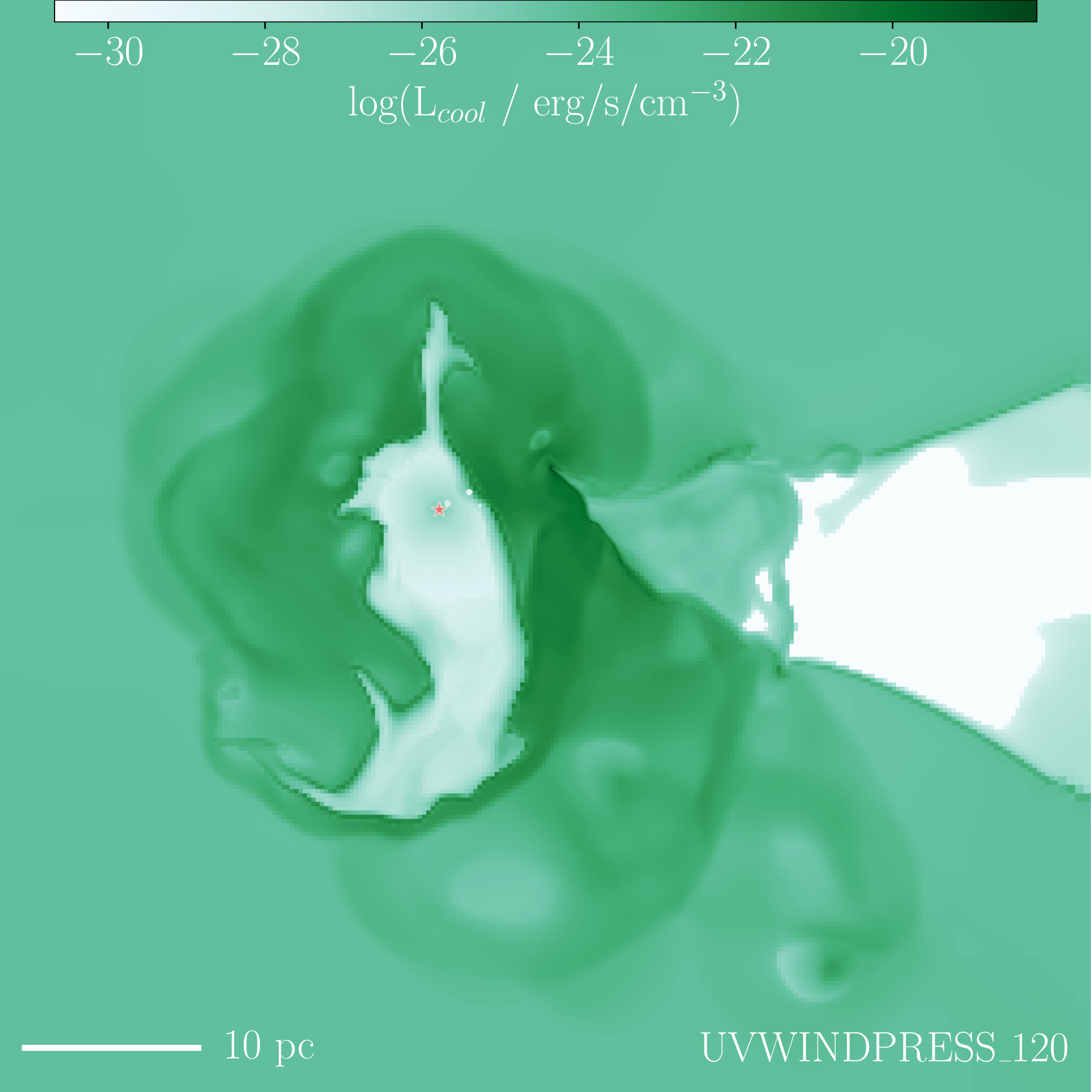}}
	\centerline{\includegraphics[width=0.80\columnwidth]{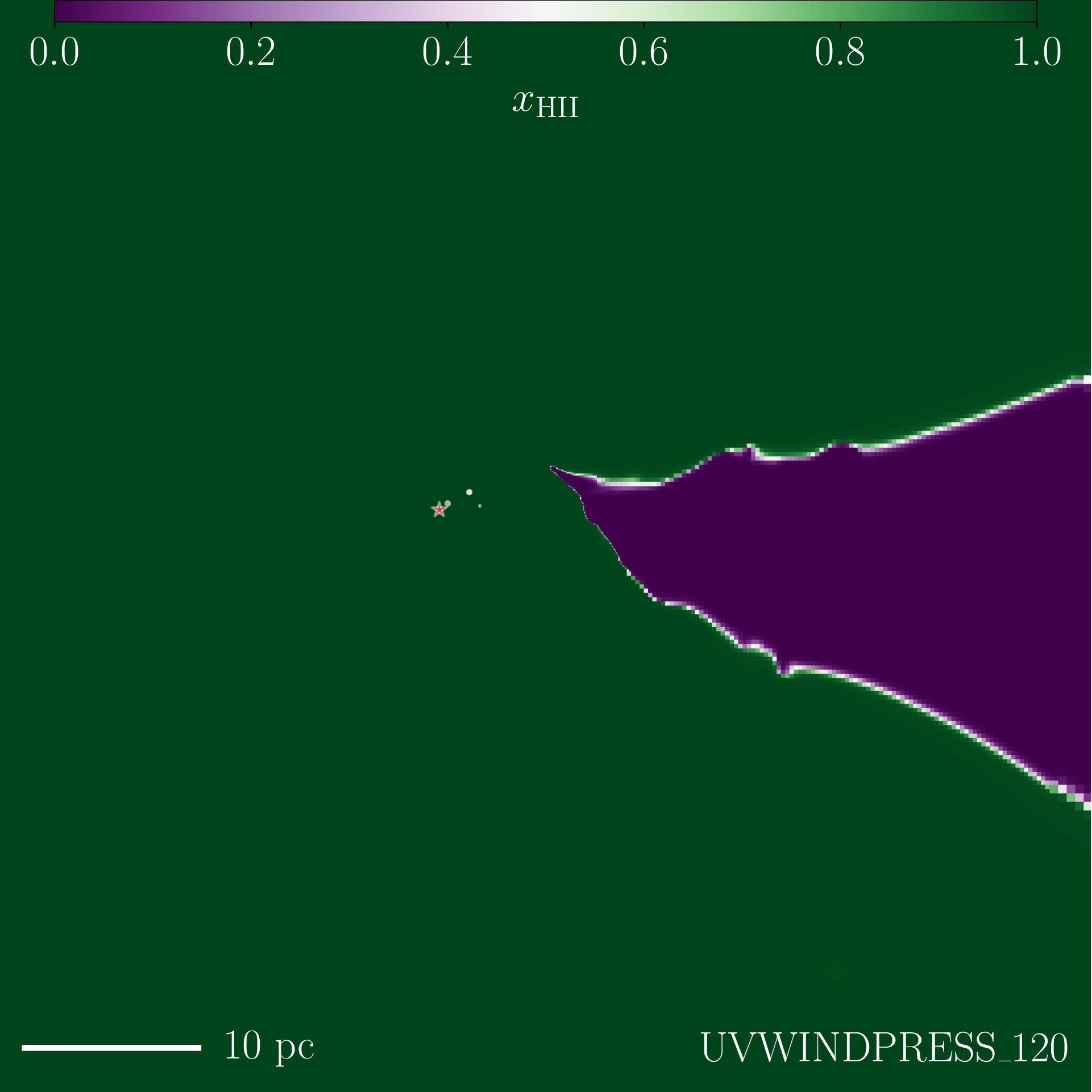}
	\includegraphics[width=0.80\columnwidth]{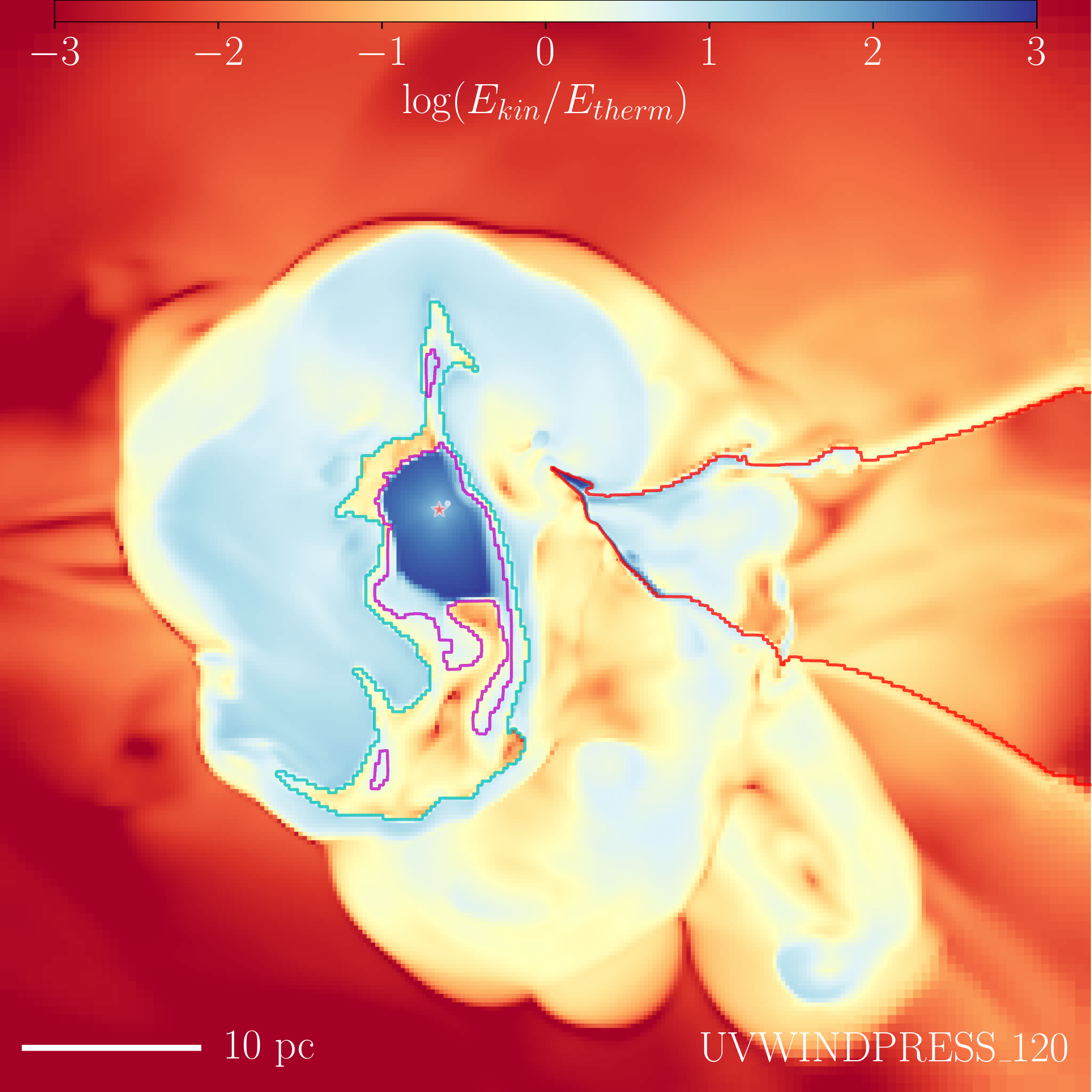}}
	\caption{Slices through the simulation volume in the UVWINDPRESS\_120 simulations at 0.4 Myr after the star is formed, showing various properties of the gas. The images are shown in the y-axis of the Cartesian volume, with the slice taken through the y-position of the star. Each image is 61 pc on-a-side, i.e. half of the total box length. From top left to bottom right, we plot mass density, temperature, thermal pressure, total radiative cooling rate (as luminosity per unit volume), hydrogen ionisation fraction and ratio of kinetic energy $E_{kin}$ to thermal energy $E_{therm}$. The radiative cooling rate is shown to illustrate where thermal energy is lost. The positions of sink particles are shown as white dots, and the star as a red star-shaped icon with a white outline. The low pressure region to the right is the shadow behind the remaining neutral gas in the cloud. In the bottom right plot we overplot contours around the wind bubble in cyan, the photoionised gas in red and gas moving faster than 1000 km/s in magenta. See Section \protect\ref{results:global} for a description of how these are defined. At 0.4 Myr in this simulation, only a small section of the neutral cloud remains.}
	\label{fig:windslices}
\end{figure*}

\subsubsection{Structure}
\label{results:evolutionwindbubble:overview}

To illustrate the typical internal structure of the wind bubble, in Figure \ref{fig:windslices} we plot slices through simulation UVWINDPRESS\_120 at 0.4 Myr after the star is formed. The slice is taken through the position of the star, from which the winds and radiation are emitted.

The structure of the bubble agrees in qualitative terms with the schematic presented in the theoretical work of \cite{Weaver1977}. In this paper, the authors describe different structures inside the wind bubble. Around the star, a free-streaming wind leaves the star as material travels outwards at the terminal velocity of the wind. This region has a lower thermal pressure than its surroundings, as seen in Figure \ref{fig:windslices}, since the flow is mostly kinetic. At some radius, the wind shocks against the ambient medium, creating a hot diffuse bubble. Since the hot wind bubble is collisionally ionised, it is almost completely transparent to ionising radiation. A warm ($10^4$ K) photoionised nebula is found outside the wind bubble.

Despite this qualitative agreement, the 3D geometry of the wind is very different from the purely spherical models of \cite{Weaver1977}. This is due to the complex 3D geometry and motions in the cloud. The wind bubble moves preferentially in certain directions of lower density out of the cloud. ``Chimneys'' of gas moving at above 1000 km/s are visible in the slice pointing along the direction the wind bubble escapes. These are shown as purple contours in the bottom right image in Figure \ref{fig:windslices}. These chimneys are not free-streaming, but are also not hydrostatic, and show evidence of bulk outwards flow.

Closer to the edge of the bubble, a mostly thermalised region of shocked gas can be seen. This part of the wind bubble is in rough thermal pressure equilibrium with the surroundings. The surrounding gas is made of ionised cloud material that has not yet expanded into the surrounding medium after being overtaken by the photoionised champagne flow.

\subsubsection{Energetics}
\label{results:energeticswindbubble}

In this section we discuss how the ability of the wind bubble to retain energy from the star determines its dynamics. We focus on two aspects of this issue. Firstly, we calculate how much energy from the stellar wind is retained in the gas, and where this energy is lost. Secondly, we look at how energy from the winds influences the dynamics of the system. This allows us to disentangle the role of winds from that of other processes in the cloud.

Throughout this section we consider two extremes for the expansion of the wind bubble, from strongest to weakest influence on the cloud, as discussed in \cite{Silich2013}. In the strongest case, the wind bubble is adiabatic and thus retains all of the energy deposited by the star as winds. It expands due to its internal overpressure as described in \cite{Avedisova1972} and \cite{Weaver1977}. In the weakest case, the bubble cools very efficiently and thus loses all of the deposited wind energy very quickly. In this mode, the bubble is accelerated only at a rate equivalent to the momentum output rate of the star. With moderate cooling rates, the bubble evolves between one of these two extremes. It is crucial to understand which best describes a given wind bubble, since both models produce drastically different outcomes for the expansion of the wind bubble, with the adiabatic mode expanding considerably faster than the efficiently cooled mode.

\begin{figure*}
\includegraphics[width=1.98\columnwidth]{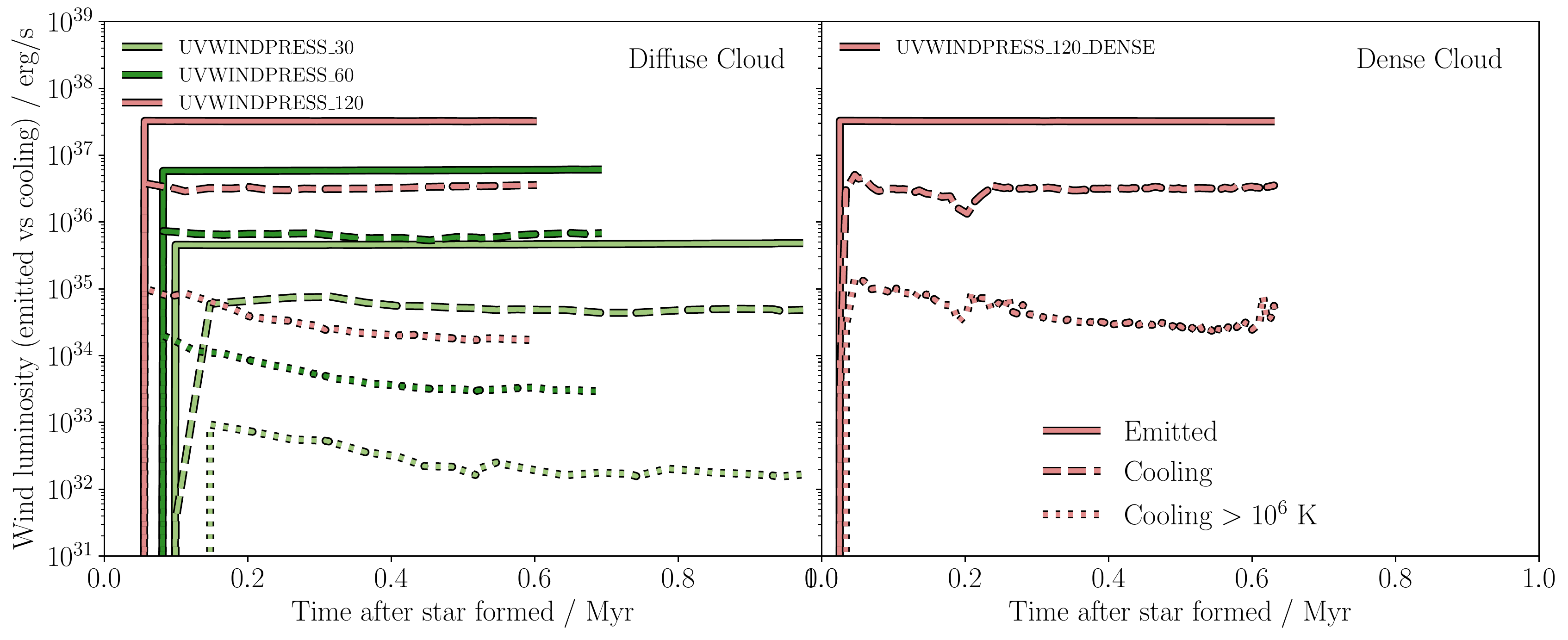}
\caption{Wind luminosity of the star at each simulation output (solid line) versus rate of energy loss to radiative cooling in the wind bubble (dashed line) and the amount of that cooling in gas above $10^6$ K (dotted line) as a function of time in each simulation containing  photoionisation, winds and radiation pressure. The wind bubble is identified as described in Section \protect\ref{results:global}. Emitted luminosity values are sampled from the stellar evolution tables (see Section \protect\ref{methods}) for the star's age in each simulation output. Cooling in gas cells above $10^6$ K is given as an upper bound on the amount of X-rays that can be emitted by the wind bubble. The gap from 0 to 0.1 Myr is due to limited frequency of simulation outputs. There is a further delay in cooling in UVWIND\_30 due to the time taken to heat the cells around the star to above $2 \times 10^4$ K. The wind bubble itself does not radiate away the majority of the energy emitted by the star as winds.}
\label{fig:windLemittedvscool}
\end{figure*}
\begin{figure}
	\includegraphics[width=0.98\columnwidth]{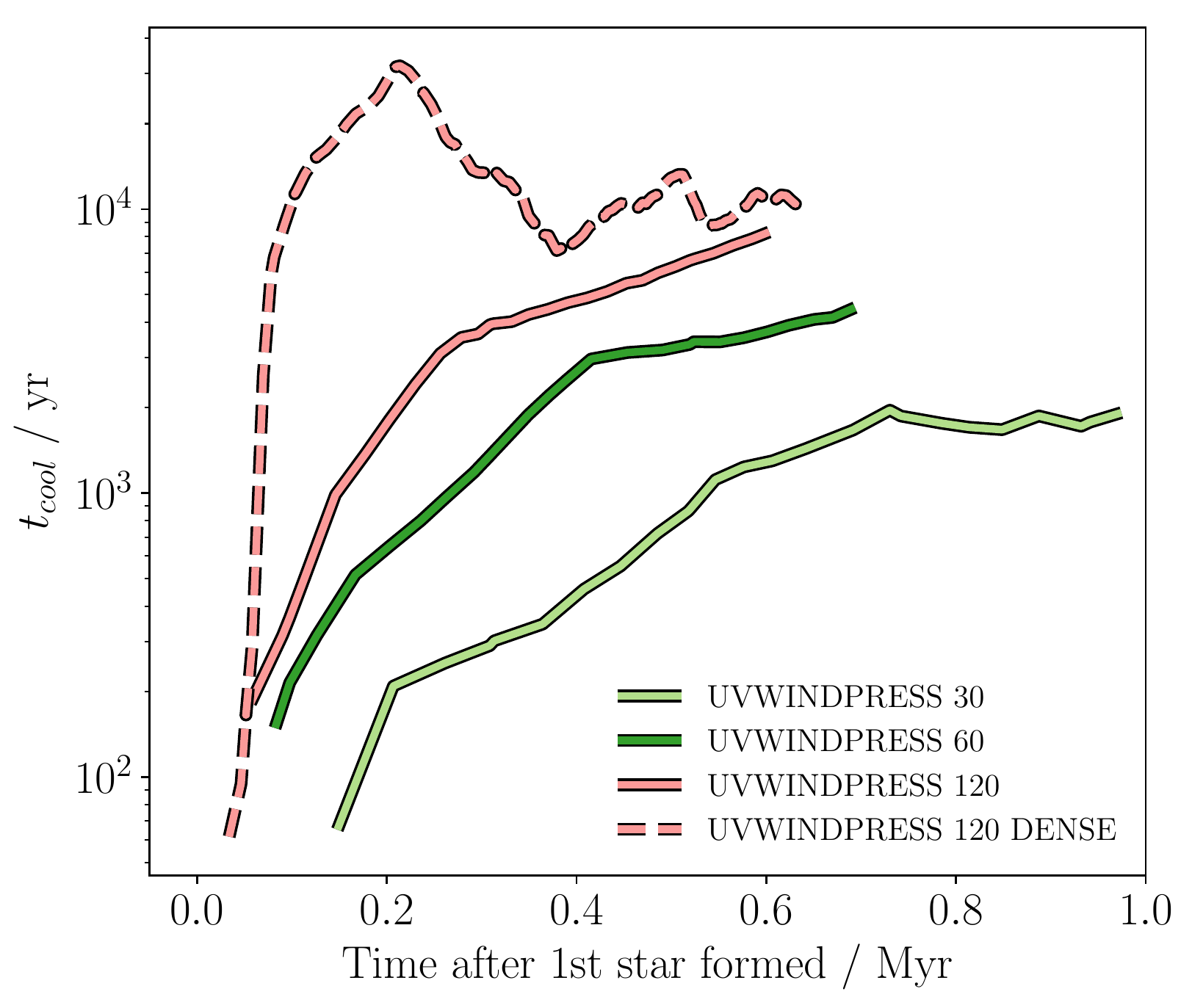}
	\caption{Estimated cooling time $t_{cool}$ of the wind bubble as a function of time in each simulation containing winds. The $t_{cool}$ is calculated using Equation \ref{equation:tcool}. The external density $n_0$ is taken to be the median density in cells immediately neighbouring the wind bubble, and the wind luminosity is taken from our stellar evolution tables. See Section \ref{results:energeticswindbubble} for a discussion of this calculation and some caveats. Lines begin where a wind bubble is first identified in an output, which depends on the output times of the simulation. $t_{cool}$ is typically between 0.1 and 1\% of the age of the star, and thus we expect the wind bubble to lose around 90-99\% of its energy.}
	\label{fig:coolingtimes}
\end{figure}

\begin{figure*}
	\includegraphics[width=1.98\columnwidth]{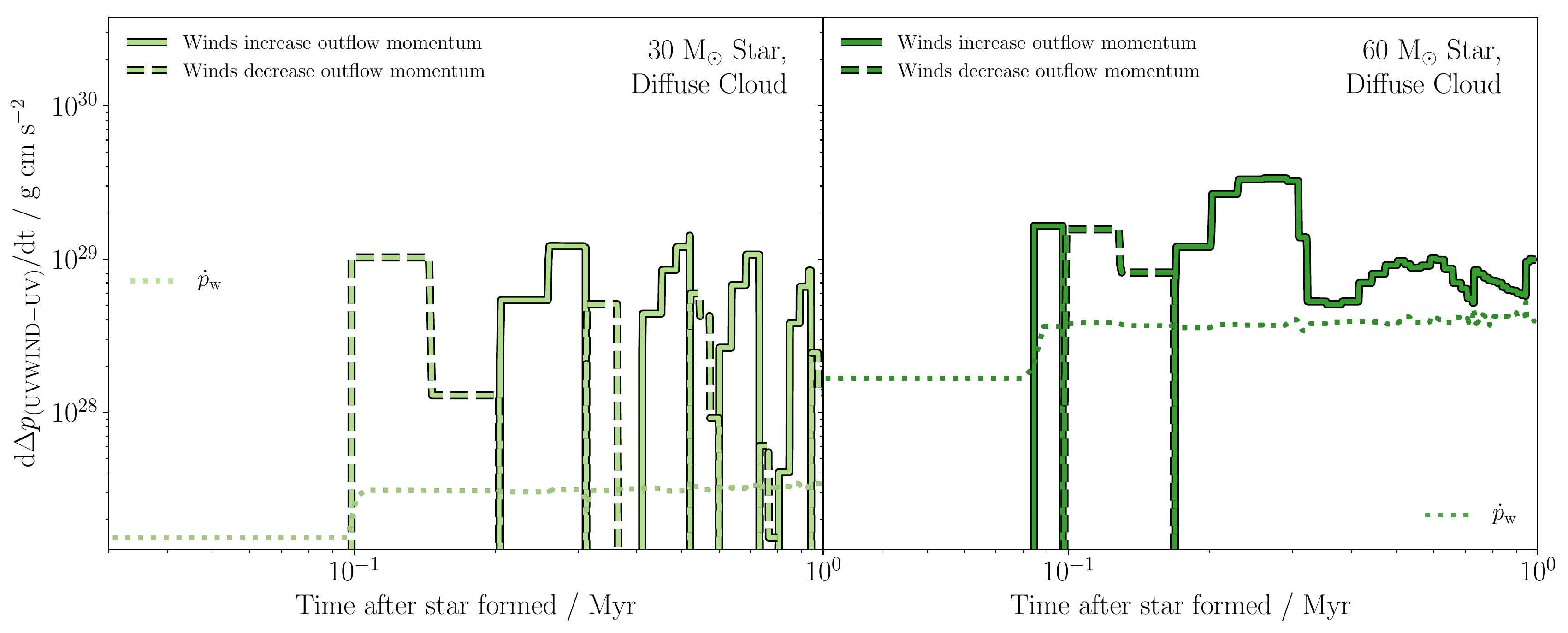}
	\includegraphics[width=1.98\columnwidth]{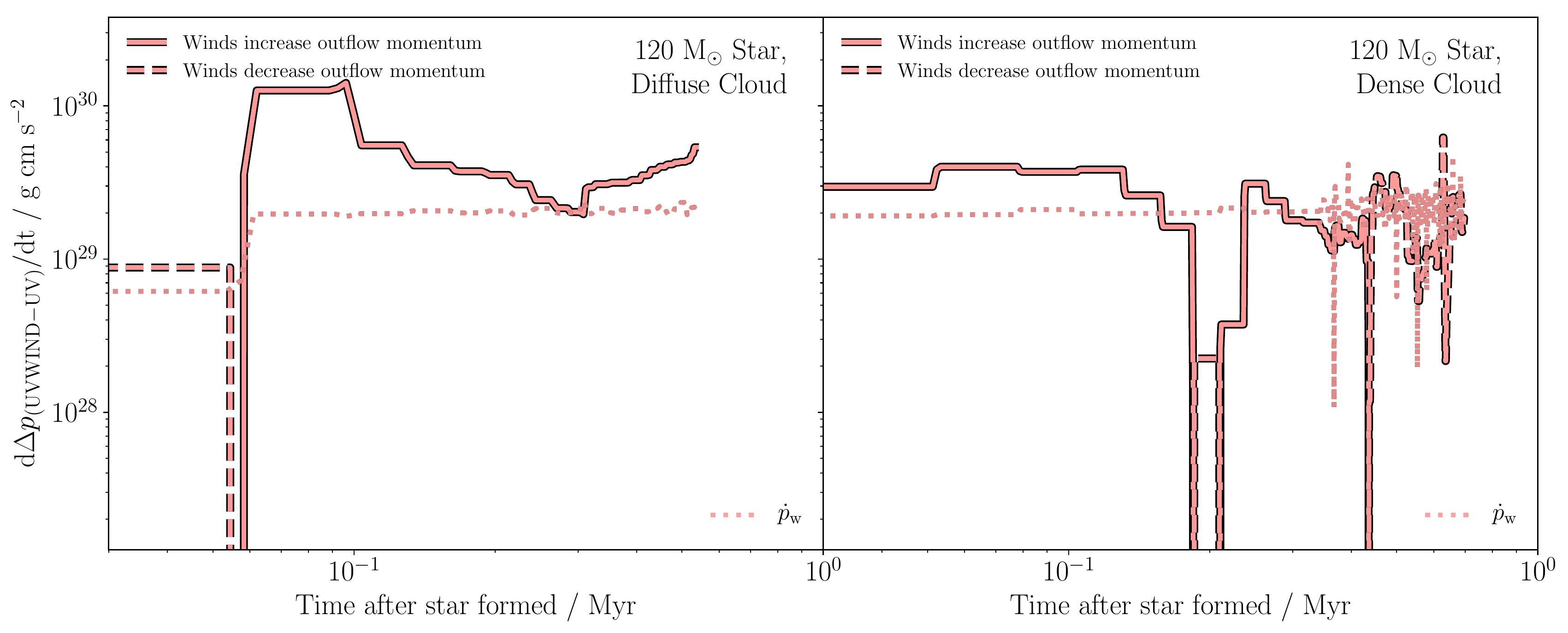}
	\caption{Figure demonstrating the influence of stellar winds on the momentum of the \protect\HII region, and the role of stored energy in the bubble. The output wind momentum deposition rate of the star $\dot{p}_{w} \equiv v_w \dot{m}_w$ is plotted as a dotted line. To directly compare to this quantity, we calculate the difference in momentum $\Delta p$ between the UVWIND and UV simulations in Figure \protect\ref{fig:momentum}. We then calculate the differential of this with time, $\mathrm{d}\Delta p_{(UVWIND-UV)}/\mathrm{d}t$. A solid line means that the UVWIND simulation has a larger increase in momentum per unit time than the UV simulation for the same star (i.e. adding winds increases momentum). A dashed line means that the UV simulation has a larger increase in momentum per unit time (i.e. a negative version of the solid line). If the solid line lies above the dotted line, it implies that the wind bubble is expanding faster than the efficiently-cooled wind bubble model described in Section \protect\ref{results:energeticswindbubble}, and is thus partially driven by pressure induced by stored energy in the wind bubble. We find that in most cases, our wind bubble follows the efficiently cooled model.}
	\label{fig:momentumrate}
\end{figure*}

In Figure \ref{fig:windLemittedvscool} we plot the energy lost to radiative cooling as a function of time in the hot wind bubble (i.e. ignoring radiative losses in the warm or cold gas) and compare it to the wind energy input from the star as a function of time. We identify cells as being in the wind bubble using the same method as in Section \ref{results:global}.  The cooling rate inside the wind bubble is at most $\sim$10\,\% of the emitted wind luminosity.  We further plot the cooling from cells above 10$^{6}$\,K, which will predominantly emit in X-rays, and find an even lower fraction of the wind bubble’s energy is lost in this gas phase. This can also be extracted from the panel displaying $L_{\rm cool}$ in Figure \ref{fig:windslices}, where it is clearly visible that the cooling rate inside the bubble is indeed low. One might thus expect the wind bubble to be close to adiabatic. However, when we measure the total energy stored in the wind bubble, it is typically only 1\,\% of the total wind energy injected. This is explained by a region of enhanced cooling along the interface between the wind bubble and the denser photoionised gas outside. In an adiabatic wind bubble solution, some of the energy in the wind bubble is stored in the dense shell around the wind bubble. In our simulations, we do not find a dense shell, and it appears that this energy is instead lost in mixing with the rapidly cooling photoionised gas.

The efficient cooling of wind bubbles through mixing with the interface between the hot wind bubble and the cooler, denser gas outside the bubble is modelled analytically in \cite{MacLow1988}. They give a cooling time $t_{cool}$ over which the wind bubble loses the majority of its energy,
\begin{equation}
t_{cool} = 2.3 \times 10^4~n_0^{-0.71} L_{38}^{0.29}~\mathrm{yr}
\label{equation:tcool}
\end{equation}
where $n_0$ is the density of the ambient medium around the wind bubble and $L_{38}$ is the wind luminosity in units of $10^{38}$ erg/s. We measure the average density just outside the wind bubble to find $n_0$, and obtain $L_{38}$ from our stellar evolution tables. We thus obtain a value of $t_{cool}$ for our simulations, which we plot in Figure \ref{fig:coolingtimes}. $t_{cool}$ is typically between 0.1\% and 1\% of the star's age. This short cooling time is consistent with our finding that the wind bubble typically retains only a small fraction of the energy injected by stellar winds ($\sim1~\%$), and that the wind energy of the star is mainly lost in the interface between the wind bubble and the photoionised gas outside it.  

\rev{It should be noted that despite the difference in cloud densities between the \textsc{dense} and \textsc{diffuse} clouds, the wind bubbles cool at a similar rate in both clouds. This can be explained by noting the geometry of the wind bubble in Figure \ref{fig:windLemittedvscool} - by 0.1 Myr, the wind bubble chimney has punched through the \textsc{dense} cloud and created a plume in the more diffuse gas around the cloud. This means that the wind bubble is less affected by mixing with the denser cloud material.}

We therefore expect our wind bubble to behave most similarly to the efficiently cooled model in \cite{Silich2013} and now compare directly to their model to determine whether this is the case.

There are two factors contributing to the pressure driving the wind bubble. One is the direct momentum injection rate from the star, $\dot{p}_w$, against the surface of the bubble. The second is the stored energy, $E_b$, acting to overpressure the bubble. The total pressure inside the bubble from both factors can be written as
\begin{equation}
P_b = \dot{p}_w / A + E_b / V ,
\end{equation}
where $A$ is the surface area of the wind bubble and $V$ is its volume. From classical mechanics, the force on the bubble's surface $F_b$ is equal to its rate of change of momentum $\dot{p}_b = P_b A$. We thus arrive at
 \begin{equation}
F_b = \dot{p}_b = \dot{p}_w + E_b (A / V) .
\label{equation:windmomentumdriving}
 \end{equation}
If $F_b$ is similar to $\dot{p}_w$, the wind bubble contains negligible stored energy, and it is driven purely by direct momentum deposition. If $F_b$ is much larger than $\dot{p}_w$, the wind bubble retains a large quantity of energy input by the star, i.e. is partially or completely adiabatic, and its expansion is driven by overpressure inside the wind bubble.
 
In Figure \ref{fig:momentumrate} we plot this comparison as a function of time. We calculate the difference in momentum between the UVWIND and UV simulations to give the contribution to the momentum from winds, assuming this is a linear perturbation to the total momentum. We then calculate the rate of change in this difference, and compare it to the momentum output rate $\dot{p}_w$ in winds from the star. From the time the wind bubble is established up to a stellar age of 0.3 Myr, winds from the 60 and $120~$\Msolar stars in the \textsc{diffuse} cloud add an order of magnitude more momentum to the wind bubble than $\dot{p}_w$, suggesting a stored quantity of energy. However, this energy is quickly lost and $\dot{p}_b$ drops to a similar value to $\dot{p}_w$. Winds in the \textsc{dense} cloud only ever expand on the level of  $\dot{p}_w$. The 30 \Msolar star does expand faster than if it were driven purely by $\dot{p}_w$, but does not expand smoothly, suggesting some influence of pressure from external forces such as turbulence. 
 
In our simulations, the wind bubble evolves according to an efficiently cooled model, driven only by the momentum from the current stellar wind output, with negligible stored energy. The high temperature inside the wind bubble is counterbalanced by its low density, and the stored energy is only around 1\% of the input wind energy.
 
 \section{Discussion}
 \label{discussion}
 
 We have so far established a picture of a wind bubble that broadly follows the classical picture in \cite{Weaver1977}, where the wind bubble expands spherically with a free-streaming volume embedded within a hot, shocked volume. However, in our simulations, the wind bubble cools rapidly and has a highly aspherical geometry that responds chaotically to structures in the photoionised cloud in which it is embedded. In this Section we compare this picture from simulations with a sample of previous analytic models and observations, and discuss how our results in Section \ref{results} should be viewed in this context.
 
\subsection{Comparison with Analytic Models}
\label{results:comparison-with-analytic-models}

\cite{Geen2019} provide algebraic expressions to describe the evolution of an efficiently cooled wind bubble embedded inside a photoionised \HII region using a hydrostatic 1D model to describe its internal structure. They characterise the behaviour of the wind bubble with a coefficient $C_w$. $C_w > 1$ if winds contribute more to the expansion of the \HII region than photoionisation, and $C_w < 1$ if photoionisation contributes more. Using Equation 26 of \cite{Geen2019}, we can write
\begin{equation}
\begin{split}
C_w = 0.0119
\left( \frac{\dot{p}_w}{10^{29}~\mathrm{g.cm/s^2}} \right)^{3/2} 
\left( \frac{Q_H}{10^{49}~\mathrm{s}^{-1}} \right)^{-3/4} \times \\  
\left( \frac{r_i}{1~\mathrm{pc}} \right)^{-3/4}
\left( \frac{c_i}{10~\mathrm{km/s}} \right)^{-3},
\end{split}
\label{wind:condition}
\end{equation}
where $\dot{p}_w$ is the momentum deposition rate from the wind, $Q_H$ is the ionising photon emission rate, $r_i$ is the radius of the ionisation front and $c_i$ is the isothermal sound speed in the photoionised gas. We set $c_i$ to 11.1 km/s, which is the sound speed in the ionised gas in our simulations. In \cite{Geen2019}, we find that $C_w > 1$ for stars above 60 \Msolar below 0.01 pc, and above 120 \Msolar below 0.1 pc.
%Using Figure 4 in \cite{Geen2019}, $c_i$ evolves slightly with stellar mass with $c_i = 8.8$ km/s for a 30 \Msolar star, $9.2$ km/s for a 60 \Msolar star and $9.7$ km/s for a 120 \Msolar star.  

Using values for our stellar sources at 1 pc, we find $C_w = 0.005$ for the 30 \Msolar star, $0.065$ for the 60 \Msolar star, and $0.28$ for the 120 \Msolar star. At larger radii, this value drops. This is broadly consistent with the simple hydrostatic model in \cite{Geen2019}, where the contribution from winds from a 30 \Msolar star is negligible, while as stellar mass increases, the contribution from winds becomes more apparent but is never the primary source of momentum from outflows on cloud scales. For more massive clusters, this may change, since the equation scales as $\dot{p}_w^2 / Q_H$. As we add more stars to the source of the \HII region, winds will play a larger role.

\cite{Geen2019} gives an equation for the ratio of the ionisation front radius $r_i$ and the wind bubble radius $r_w$
\begin{equation}
\left (\frac{r_i}{r_w}  \right )^{4} = 2^{1/3} C_w^{-4/3} + \frac{r_i}{r_w} .
\label{wind:radiusratio}
\end{equation}

\begin{figure*}
	\includegraphics[width=1.98\columnwidth]{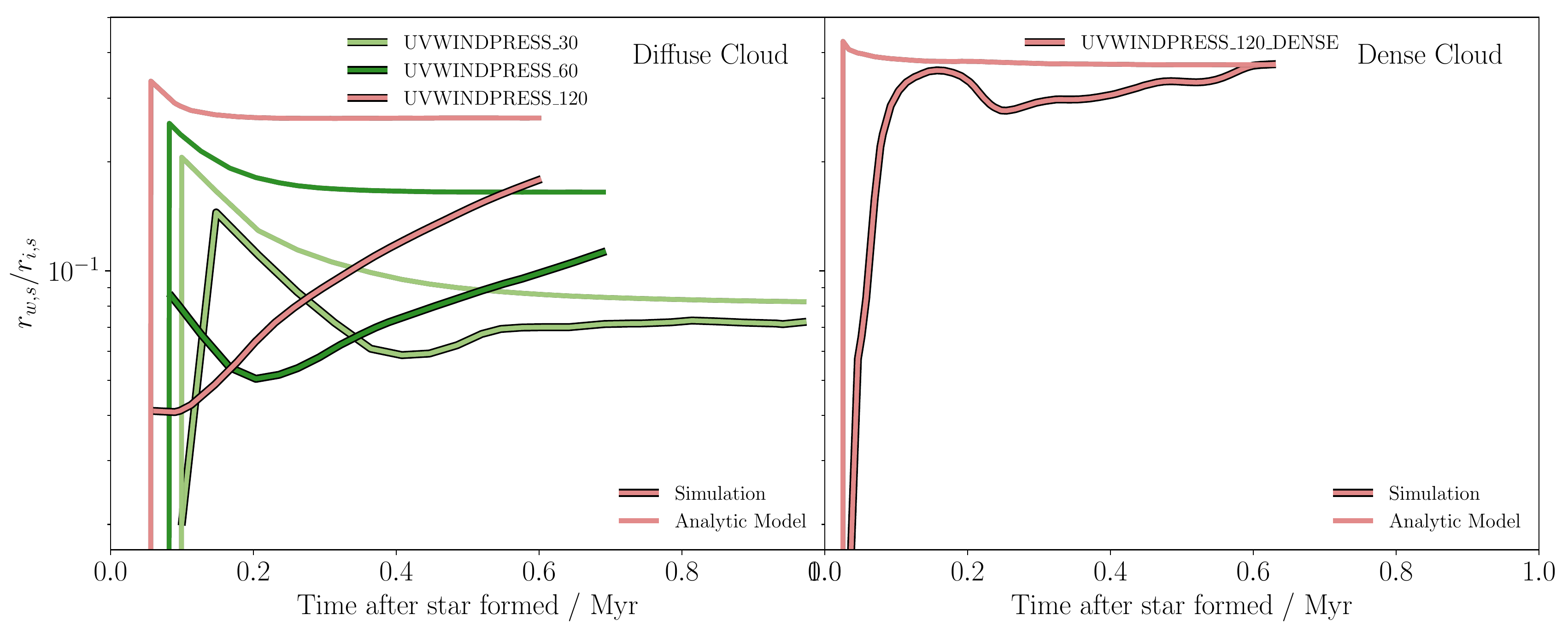}
	\caption{Ratio between radius of wind bubble and radius of ionised region. The thick outlined solid lines show results for our simulations, i.e. ($V_w / V_i$)$^{1/3}$, where $V_w$ is the volume of the wind bubble and $V_i$ is the volume of ionised gas. The thin solid lines show the results of Equation \ref{wind:radiusratio}.}
	\label{fig:windradiusratio}
\end{figure*}

In Figure \ref{fig:windradiusratio}, we compare this expression to the relative radii of the wind bubble and ionised nebula over time in each simulation with winds. We again take the sphericised radii $r_{i,s}$ and $r_{w,s}$ shown in Figure \ref{fig:radius} as representative radii, where $r_{w,s} / r_{i,s} \equiv (V_w / V_i)^{1/3}$. $V_w$ is the volume of the wind bubble and $V_i$ is the volume of ionised gas.

In general, the analytic model in Equation \ref{wind:radiusratio} somewhat overestimates of the radius of the wind bubble at early times. At late times, the analytic equation matches the simulations more closely. Note that once the \HII region reaches the edge of the box, the simulation is unable to track its whole evolution.

One difference between the analytic model and our simulations is the geometry of the wind bubble. An elongated wind bubble has a higher surface area to volume ratio than a sphere. This means that the highly aspherical wind bubbles in our simulations will, for a given internal pressure, typically occupy a smaller volume than they would if they were purely spherical. The smaller, trapped wind bubble around the 30 \Msolar star provides a better match to the analytic model.

A second difference is that the spherical 1D analytic model misses 3D dense gas clumps and filaments in the cloud that follow the turbulent gas flows in the cloud. The effect of this is seen in the \textsc{dense} cloud, where the wind radius drops temporarily as the chimney structure is cut off and the hot plume cools. However, in this simulation the analytic model is otherwise a reasonable match to the simulation.

Our results agree qualitatively with other analytic models of the interaction between winds and photoionisation. \cite{Capriotti2001} create a set of analytic models taking into account the cooling rate of the wind bubble, while \cite{Haid2018} produce a model that assumes zero cooling to compare to their simulations. In both cases, photoionisation remains the main driver of the \HII region except in cases where the medium is already ionised, such as the diffuse interstellar medium.

\cite{Dale2014} produce a model that assumes the wind bubble is efficiently cooled, as in the \cite{Geen2019} model. However, they use a \cite{SpitzerLyman1978} model for the expansion of the \HII region, which requires a uniform background gas density. \cite{Dale2014} find that their model underpredicts $r_w/r_i$, which they argue is due to leakage of the \HII region into the gas outside the cloud. The \cite{Geen2019} model uses a power law density field $\rho \propto r^{-2}$, which captures some of this behaviour and produces a closer fit between the simulation results and analytic model. Note that the \cite{Dale2014} simulations include multiple wind and radiation sources, which adds extra complications to the comparison.

The picture thus far is one in which winds are a secondary effect in the expansion of \HII regions and the destruction of molecular clouds. They produce complex, chaotic structures, but these follow the structures shaped by other processes rather than setting the conditions in the cloud themselves.

\subsection{Comparison with Observations}
\label{results:observations}

\begin{figure*}
	\centerline{\includegraphics[width=0.99\columnwidth]{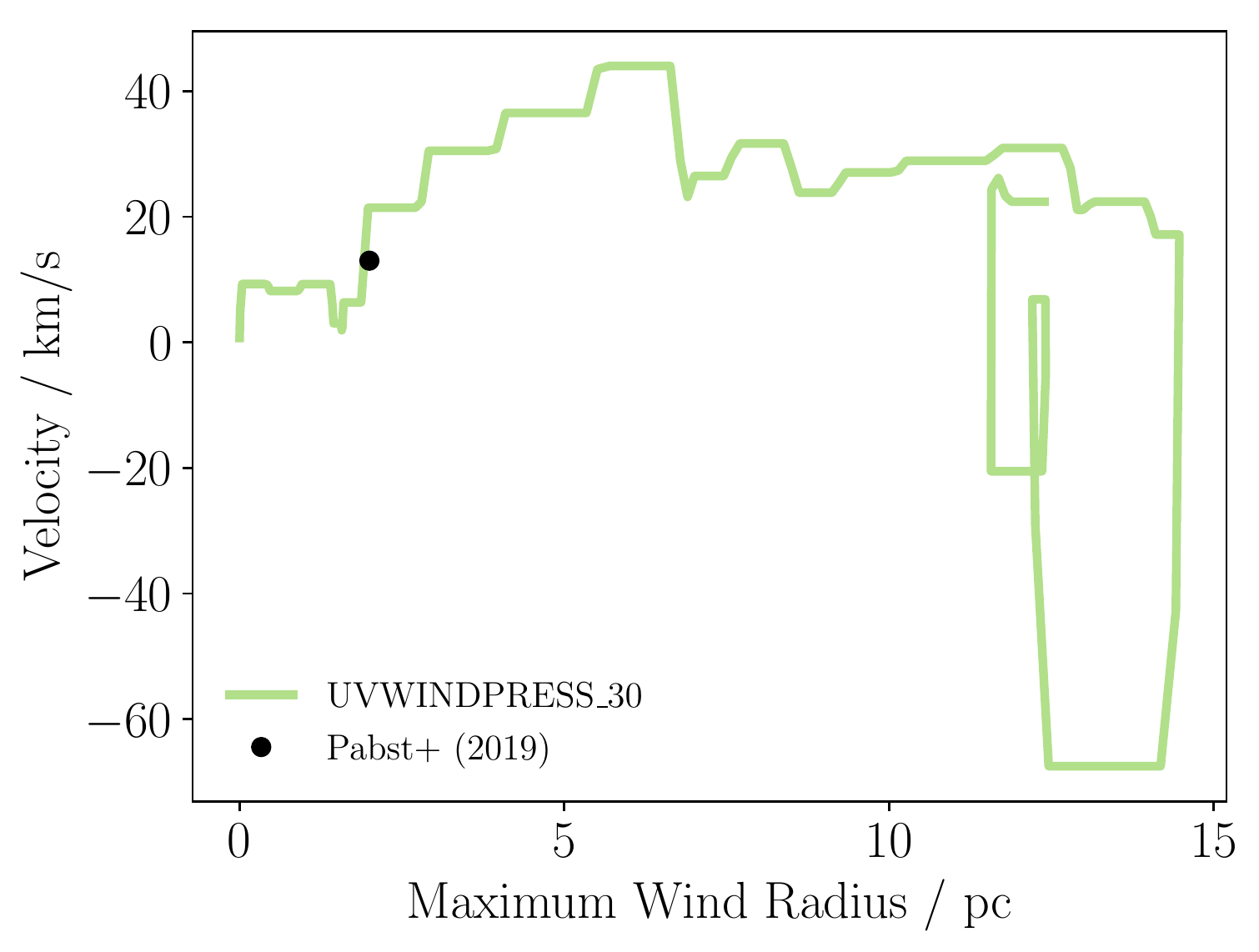} \includegraphics[width=0.99\columnwidth]{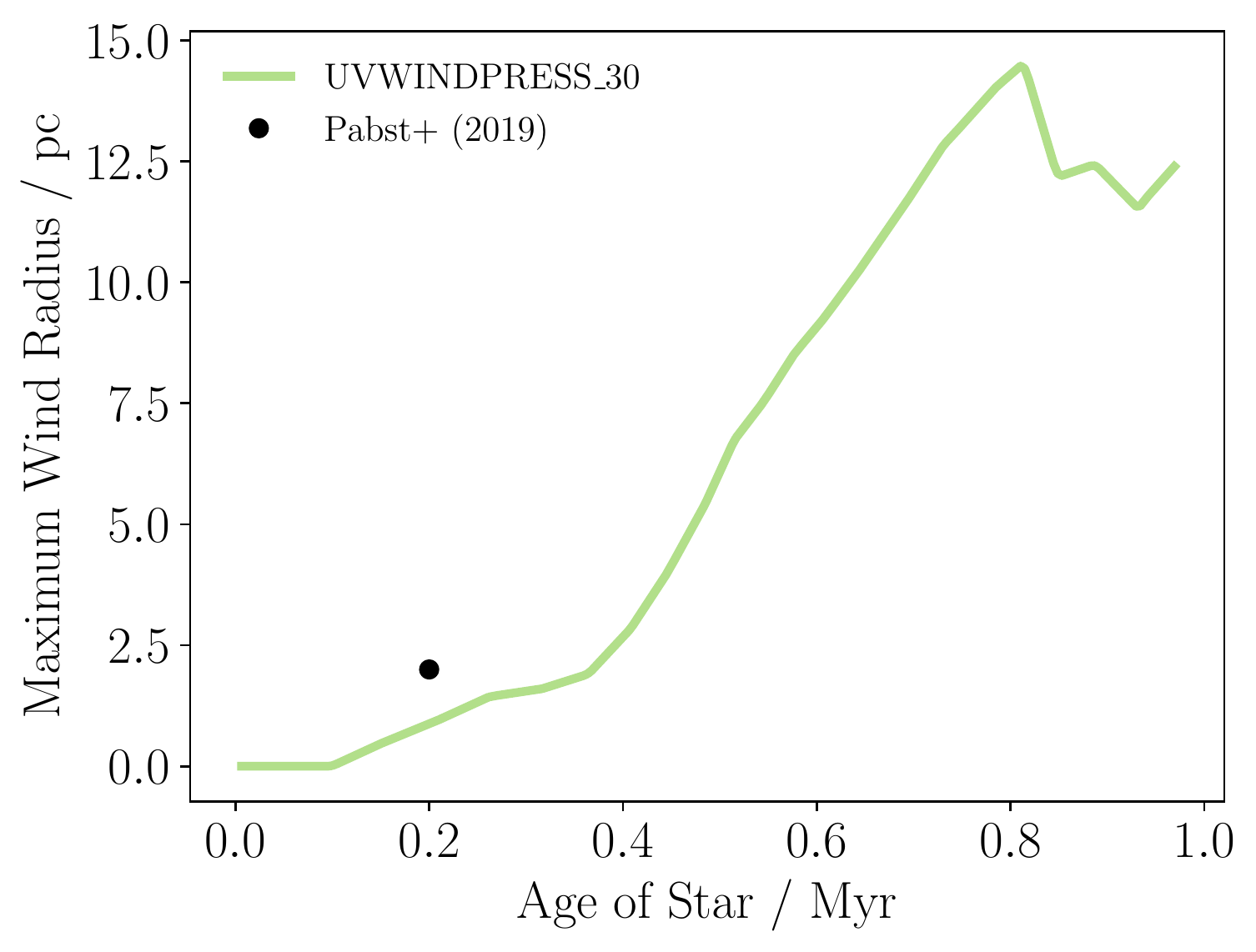}}
	\caption{Evolution of the wind bubble's maximum radius from the star, $r_{w,max}$ and the expansion velocity of the wind bubble $v_{w,max}$ in UVWINDPRESS30, the simulation most closely matching $\theta^1$ Ori C \protect\citep{Balega2014} and its host cloud \protect\citep{Geen2017}. $v_{w,max}$ is defined as the rate of change in $r_{w,max}$. Left: $v_{w,max}$ versus $r_{w,max}$. Right: $r_{w,max}$ versus the age of the star. The observationally-derived results of \protect\cite{Pabst2019} are overlaid as a point. The maximum radius of the wind bubble varies non-linearly significantly with time, due to the complex geometry and behaviour of the wind bubble. Our results overlap the velocity and radius measurement of \protect\cite{Pabst2019}.}
	\label{fig:maxradiusplots}
\end{figure*}

\begin{figure*}
\centerline{\includegraphics[width=0.66\columnwidth]{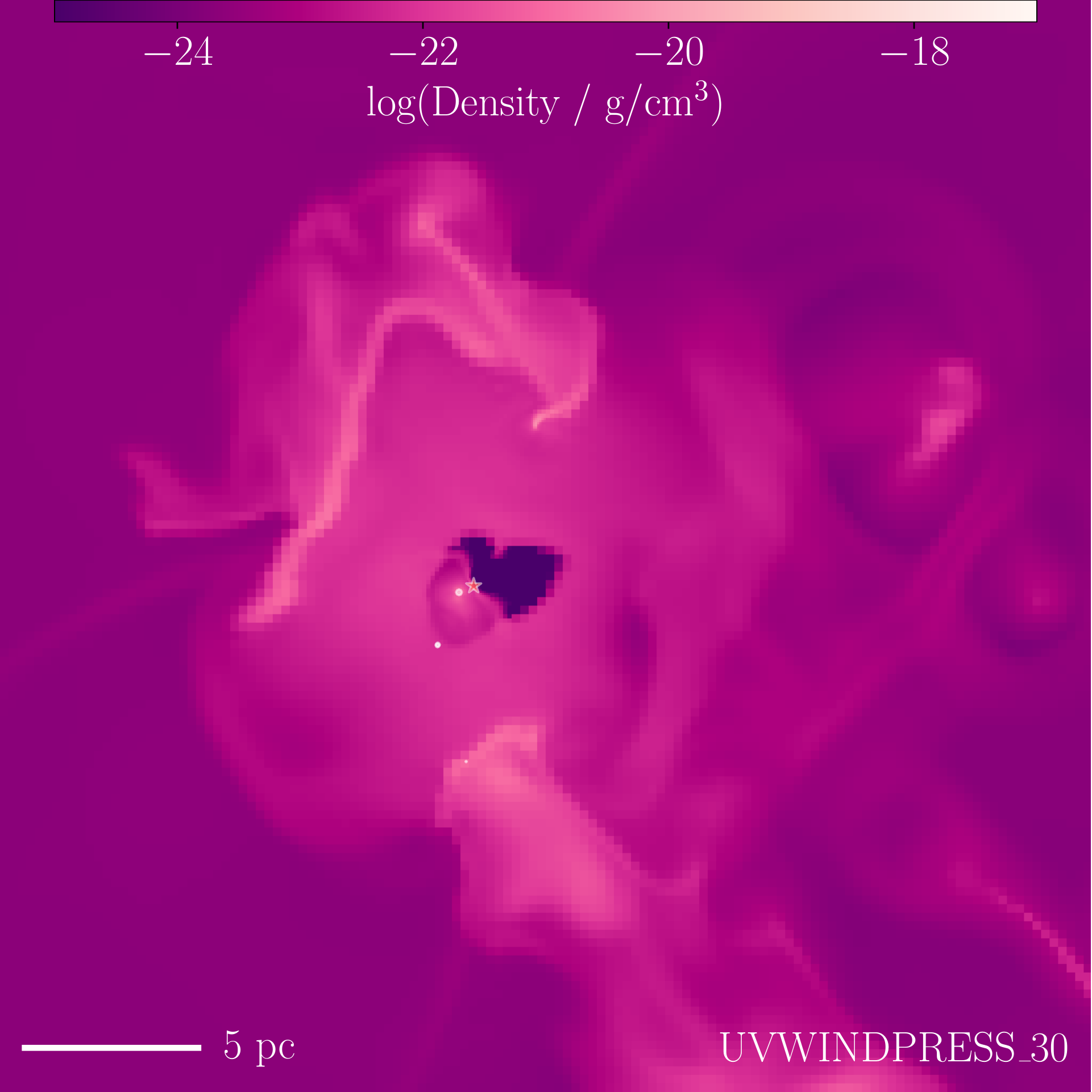} \includegraphics[width=0.66\columnwidth]{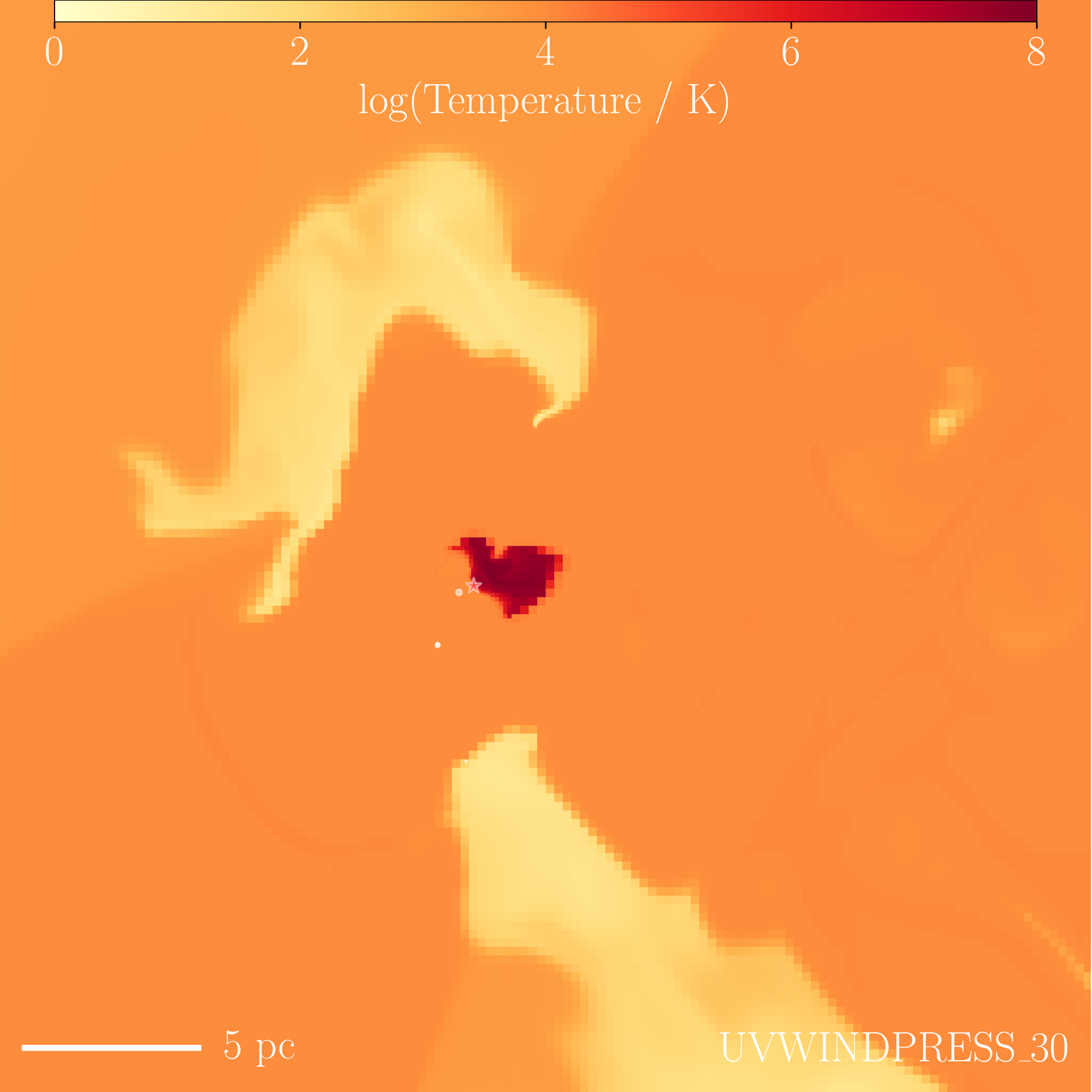}
\includegraphics[width=0.66\columnwidth]{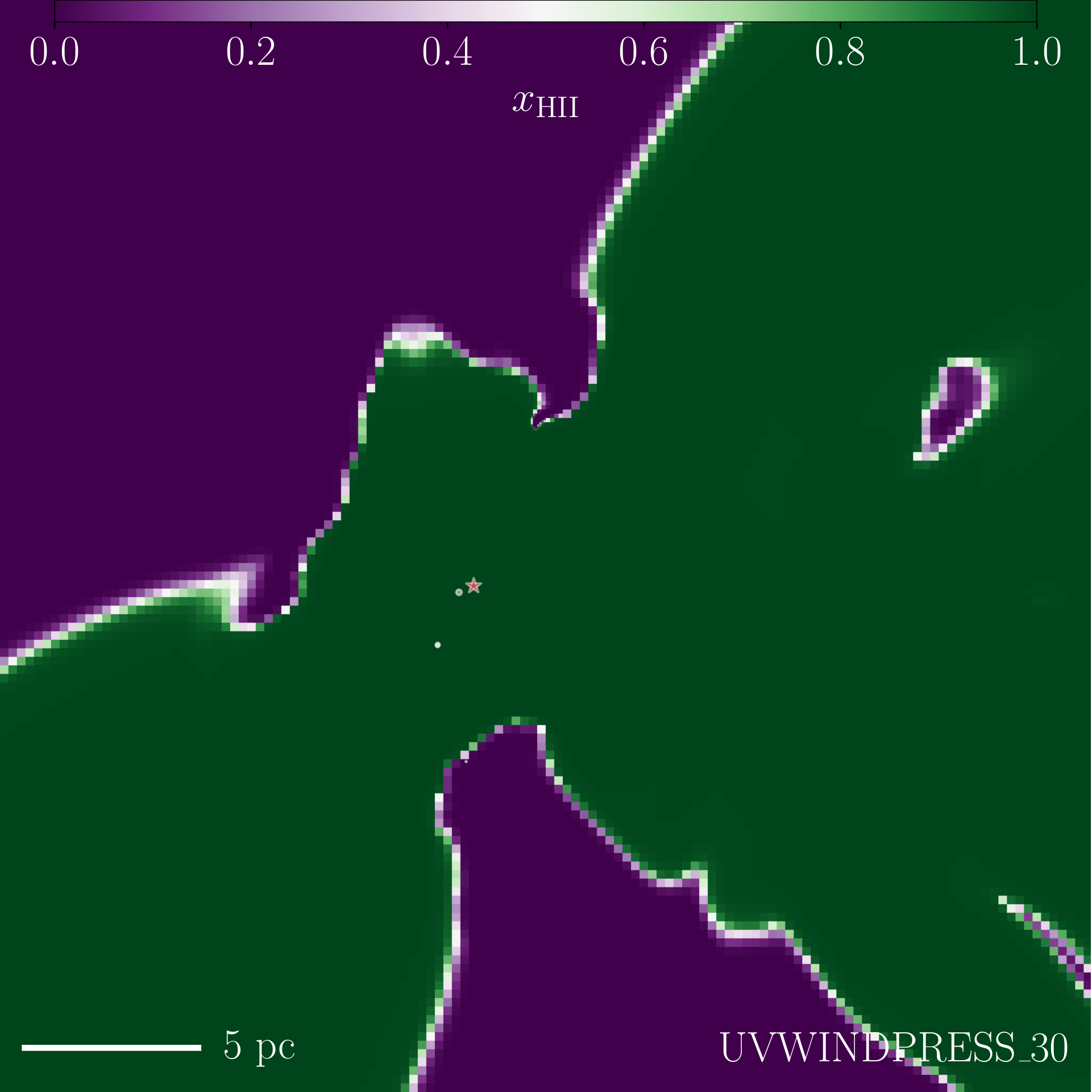}}
\caption{Slices through the position of the star in the simulation containing a 30 \Msolar star with UV photoionisation, radiation pressure and winds showing, from left to right, the gas density, temperature and ionisation state. There is no sign of a dense shell around the wind bubble, and ionising radiation escapes from the cloud. This is in contrast to the Orion Veil nebula, which has a dense shell containing neutral hydrogen.}
\label{fig:slices30Msun}
\end{figure*}

In this Section we confront our simulations with observations of the Orion nebula by \cite{Pabst2019}. A subset of our simulations are to some extent representative of the observed Orion nebula. The \textsc{diffuse} cloud in this paper was chosen to match the global density distribution of the nearby Gould belt clouds, including Orion \cite[see][]{Geen2017}. \rev{These cloud conditions are typical for the Milky Way. Density, metallicity and ISM pressure can vary for extragalactic environments \citep[e.g.][]{Gurvich2020}}. The Orion Veil nebula is driven largely by a single young massive star, $\theta^1$ Ori C, which has a mass of 33 \Msolar \citep{Balega2014}, with error bars of 5 \Msolar that cover the 30 \Msolar star we simulate.

The conclusions of \cite{Pabst2019} are that the Orion Veil nebula is a wind-driven bubble that cools inefficiently with no sign of influence from photoionisation feedback. This is at odds with our conclusions, and the conclusions of other theoretical works. We thus confront our simulations with the \cite{Pabst2019} observations of Orion to try to determine possible reasons for this discrepancy.

\subsubsection{Bulk Wind Bubble Properties}

We first compare the radius, velocity and age of the wind bubble to the values in \cite{Pabst2019}. We plot these values in Figure \ref{fig:maxradiusplots}. We use the maximum radius of the wind bubble, since the Orion Veil nebula expands in only one direction away from the star, constrained by dense gas in the opposite direction. See \cite{Pellegrini2007} for a discussion of the constrained part of the nebula. The maximum radius is a lot less stable than the spherically-averaged radius shown in Figure \ref{fig:radius}, since the geometry of the wind bubble is chaotic. The velocity evolution is particularly chaotic, since wind bubbles can grow and collapse rapidly in certain directions.

The results of \cite{Pabst2019} sit on top of our results in velocity-radius space. Our bubble reaches this radius at a later time than the predictions of \cite{Pabst2019}, approx 0.3 to 0.4 Myr versus 0.2 Myr. However, since the age in \cite{Pabst2019} is estimated using a simple \cite{Weaver1977} analytic calculation, and our initial conditions may vary appreciably from those in Orion, it is not unreasonable to expect different ages for the wind bubble. They also argue for a negligible role from radiation pressure, as do we.

\cite{Pabst2019} estimate a shell mass of $2600~$\Msolar, with a lower bound at $900~$\Msolar and upper bound at $3400~$\Msolar depending on assumptions in how it is calculated. For comparison, our \textsc{dense} cloud contains roughly 1000 \Msolar in the 2 pc radius around the star, which is close to the lower bound of the observed estimate. Assuming the observed mass estimate is correct, this suggests that the volume immediately around $\theta^1$ Ori C is relatively dense.

In Figure \ref{fig:slices30Msun} we show slices through the region. The geometry of the wind bubble is roughly similar to the Orion Veil nebula, i.e. a hemisphere bounded on one side by denser gas. The gas on the other side of the nebula is already photoionised, unlike the Orion Veil nebula. However, we do not expect a perfect match between the simulations and observations in terms of geometry, since the system is turbulent and chaotic. We do not run several realisations of the cloud with different initial seeds to test the effect of chaos on the system as we do in \cite{Geen2018} for reasons of computational cost.

\cite{Pabst2019} find that the X-ray emission rate of the bubble is approximately $4\times10^{31}~$erg / s. This is roughly an order of magnitude lower than the cooling rate of hot gas in Figure \ref{fig:windLemittedvscool}, although they claim that many X-rays in Orion are absorbed by intervening gas in the denser lines of sight. \cite{Guedel2007} also argue that Orion appears to be mostly filled with wind-shocked gas using X-ray observations. They argue that regions of the Orion Veil nebula that do not appear to emit X-rays are surrounded by denser gas that absorbs these photons. This is plausible, although as we show, the geometry of wind bubbles can be highly aspherical and it is possible for the wind bubble to reach large radii while not filling volumes closer to the star.

In Section \ref{results:energeticswindbubble}, we find the energy injected by stellar winds is mostly lost to radiative cooling, and so the wind bubble follows a model in which its expansion is purely driven by the momentum deposition from the star. By integrating the momentum over the star's lifetime, we can determine whether the wind bubble follows this path or retains a significant amount of energy. The total momentum deposited by the 30 \Msolar star in winds after 0.2 Myr is around $3 \times 10^{40}~$g$~$cm/s. Using the observed shell mass of $M_s=2600~$\Msolar at 0.2 Myr travelling at 13 km/s, we get a shell momentum of $7 \times 10^{42}$ g cm/s. This is much higher than the direct momentum injection from the star. Using the lower shell mass estimate $M_s=900~$\Msolar also gives a much higher momentum. Assuming that the Orion Veil nebula is driven principally by winds as \cite{Pabst2019} suggest, and using Equation \ref{equation:windmomentumdriving}, this implies that the Orion Veil nebula is driven by a store of thermal energy in the wind bubble, rather than being completely cooled. 

\subsubsection{Trapping of Ionising Photons}

The observed Orion Veil nebula appears to have a large, neutral shell around the wind bubble \citep{vanderWerf2013}. By comparison, the wind bubbles in our simulations sit inside a larger photoionised \HII region, with little or no sign of a denser shell around the wind bubble (see Figure \ref{fig:slices30Msun}). We discuss briefly why we do not find this phenomenon in our simulation results.

We take the observational results for the shell, $M_s=2600~$\Msolar at $r_s=2~$pc, moving at $v_s=13~$km/s into a medium with hydrogen number density $n_0=1400~$cm$^{-3}$. Using equation 67 in \cite{Weaver1977} and making the simplifying assumption that this is a spherical shell, we can estimate the hydrogen number density of the shell
\begin{equation}
n_s = n_0 (v_s / c_0)^2
\end{equation}
where $c_0$ is the sound speed in the neutral gas outside, and which we approximate to be 1 km/s. Using the values given above, we find $n_s \simeq 2.4\times10^5$ cm$^{-3}$. The volume of the shell $V_s$ is given by $4 \pi r_s^2 \Delta r$, where $\Delta r$ is the shell thickness. Approximating $M_s / V_s = n_s m_H/X$, we find $\Delta r\simeq 0.007~$pc, which is 4-5 times smaller than our finest grid cell size. 

The total recombination rate of such a shell is $10^{53}$ photons/s, or $10^4$ times the photon emission rate of the star. In other words, the shell should easily absorb all ionising photons coming from the star. If we assume that the shell around the wind bubble can only reach a thickness of 0.1 pc, or $\sim3$ cells at our highest level of refinement, the total photon absorption rate is still $\sim10^{52}$ photons/s, which is higher than the photon emission rate from the star. In principle, with sufficient adaptive refinement, our simulations should be able to capture the trapping of ionising photons by the shell.

As we noted earlier, a shell mass of $M_s=2600~$\Msolar implies that the background density around $\theta^1$ Ori C is significantly higher than we find in our simulations. This may explain why such a dense shell is produced, and hence why the ionising photons are trapped. It also raises the question of why the wind bubble does not cool more efficiently, however, as modelled in Equation \ref{equation:tcool}. More targeted simulations that make direct comparisons to specific observations are needed to understand why the case of Orion appears to differ from theoretical predictions to date about how photoionisation and winds should interact.

\subsection{Further Considerations}
\label{effective:intheend}

%The results of the simulations in this work agree with previous studies using numerical simulations and analytic theory. There is some agreement with comparable observed regions, although more careful study is needed to properly understand the behaviour of these observed systems.

Full 3D hydrodynamic simulations are still relatively expensive due to the cost of simulating fast, hot flows such as stellar winds. This presents a problem for exploring a larger parameter space. By reducing the problem to 1D and making certain simplifications, some of these limitations can be overcome. Recent 1D analytic models by, e.g., \cite{Rahner2017} and \cite{Pellegrini2019} are able to match certain observed properties of nearby \HII regions around massive clusters.

There is still a need for 3D simulations to capture the full behaviour of molecular clouds with embedded stellar wind bubbles. We have shown that the behaviour of wind bubbles around single stellar sources is already complex. Various other authors have already begun to explore the interactions between wind bubbles, which adds a further layer of complication. \cite{Rogers2013} and \cite{Dale2014} perform simulations of multiple massive stellar wind sources, finding that the ablation of dense clumps as multiple wind bubbles merge around them provides an additional cooling channel. Clumping and shell fragmentation also occurs in the interaction between free-streaming winds on small scales, such as around close binaries \citep{Calderon2020} and the Galactic centre \citep{Calderon2020a}. A further channel for clumping is the time-variability of wind velocities in the Wolf-Rayet phase \citep[see review by][]{Wade2012}.

Cooling rates from wind bubbles are also a matter of debate. Our simulations match the analytic model of \cite{MacLow1988}, in which material evaporated from the wind bubble shell mixes with the hot gas and causes efficient cooling. However, a more detailed understanding of the microphysics of the shell is needed to determine whether this happens in all cases. As we resolve smaller scales, thermal conduction and thermal instabilities become important \citep{Koyama2004}. Authors such as \cite{Gentry2016} argue for lower cooling rates in hot superwind bubbles formed by multiple supernovae. Cosmic rays can also retain some dynamically important energy in hot interstellar bubbles, as they can interact with gas at larger radii \citep[e.g.][]{Wadepuhl2011,Dashyan2020}. The microphysics of gas cooling is thus important in understanding the dynamics of hot bubbles such as those driven by stellar winds.

The stellar evolution framework described in this work can be extended to cover longer timescales and larger spatial scales. In this regime, the late stage behaviour of massive stars becomes more important. As wind bubbles expand and merge into superbubbles, winds are boosted at late times relative to ionising radiation emission, as found in \cite{Rahner2017}. However, this depends on the stellar evolution model used. \cite{Sana2012} observed that the majority of massive stars are in interacting binaries, strongly affecting their evolution. One of the results of this is that the extended envelopes of these stars can be removed by binary interactions, allowing for higher ionising emission rates at late times as in \cite{Gotberg2019}.

\section{Conclusions}
\label{conclusions}

We simulate a set of molecular clouds containing a single massive star of either 30, 60 or 120 \Msolar formed self-consistently using sink particles in clouds of two different densities. We track the expansion of \HII regions due to over pressure caused by photoionisation and radiation pressure from photons and stellar winds produced by the star.

We find that winds contribute at most 10\% to the outflow momentum in the first Myr of the lifetime of the star, and have only a small impact on the radial expansion of the \HII region. The contribution from winds in our simulations shows limited or no evidence for large quantities of stored energy driving expansion, as expected in models of adiabatic wind bubbles. Radiation pressure has a negligible effect on the evolution of the systems modelled in this paper.

While the volume and momentum of the simulated wind bubbles evolves smoothly, the geometry of wind bubbles is highly aspherical and chaotic. The high characteristic velocity of wind bubbles means that they can rapidly evolve to fill pressure gradients in clouds and \HII regions. Outside of the classical free-streaming radius, the structure of these wind bubbles is characterised by kinetic-energy-driven ``chimneys'' and thermally-driven ``plumes''. These plumes can be cut off by changes in the denser gas flows in the cloud and \HII region, in some cases leading to hot bubbles not connected to a stellar source.

Our simulations provide good agreement to previous simulations and analytic models in key aspects while demonstrating the need for full 3D simulations to capture the complex behaviour of stellar winds. Comparison to the Orion Veil nebula match certain bulk properties, but differences between the two systems suggest that new simulations designed to match the specific environment of observed regions are needed to close the gap between observations and theory.

\section{Data Availability}
\label{data-management}

This paper has been prepared according to the Research Data Management plan of the Anton Pannekoek Institute for Astronomy at the University of Amsterdam. Details of the data products and scripts used to generate the figures in this paper are found via DOI reference \textit{10.5281/zenodo.3696806}.

\section*{Acknowledgements}

The authors would like to thank Jo Puls, Xander Tielens, Cornelia Pabst, Patrick Hennebelle, Eric Pellegrini, Daniel Rahner, Ralf Klessen, Lex Kaper and Zsolt Keszthelyi for useful discussions during the writing of this paper. \rev{The authors would like to further thank the anonymous referee for their careful and insightful comments that helped improve the clarity of the manuscript.} The simulations in this paper were performed on the Dutch National Supercomputing cluster Cartesius at SURFsara. The authors gratefully acknowledge the data storage service SDS@hd supported by the Ministry of Science, Research and the Arts Baden-W\"urttemberg (MWK) and the German Research Foundation (DFG) through grant INST 35/1314-1 FUGG. Postprocessing was performed on the ISMSIM computing resources hosted by ITA, ZAH, University of Heidelberg. The project also made use of the SURFsara ResearchDrive facility for remote data sharing. SG acknowledges support from a NOVA grant for the theory of massive star formation.

%%%%%%%%%%%%%%%%%%%%%%%%%%%%%%%%%%%%%%%%%%%%%%%%%%



\bibliographystyle{mnras}
\bibliography{samgeen} % if your bibtex file is called example.bib

%%%%%%%%%%%%%%%%%%%%%%%%%%%%%%%%%%%%%%%%%%%%%%%%%%

%%%%%%%%%%%%%%%%% APPENDICES %%%%%%%%%%%%%%%%%%%%%

\appendix
%
%\section{Appendix Blah}
%
%APPENDIX

\section{Stellar Tracks}
\label{appendix:stellartracks}

\begin{figure*}
	\includegraphics[width=1\columnwidth]{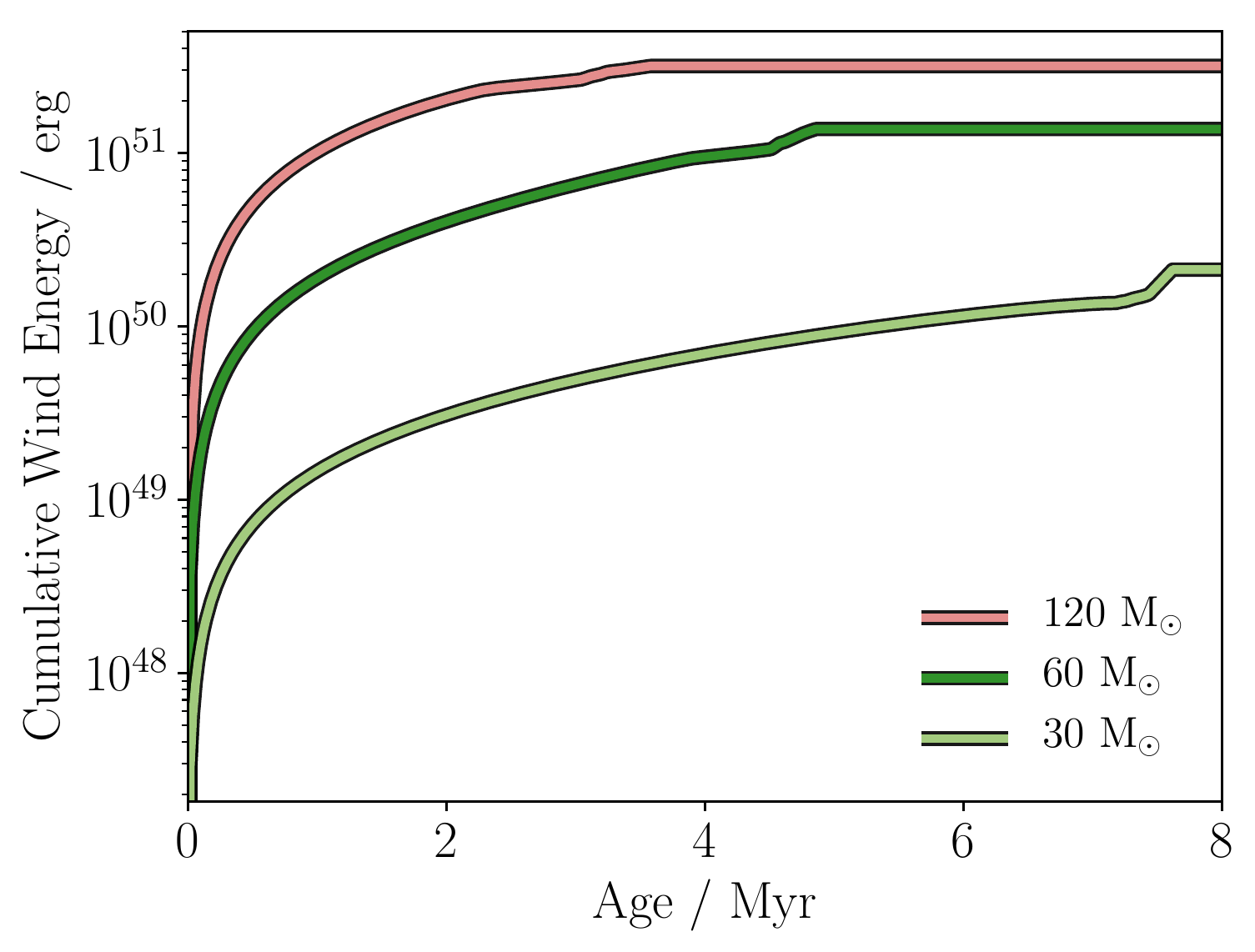} 
	\includegraphics[width=1\columnwidth]{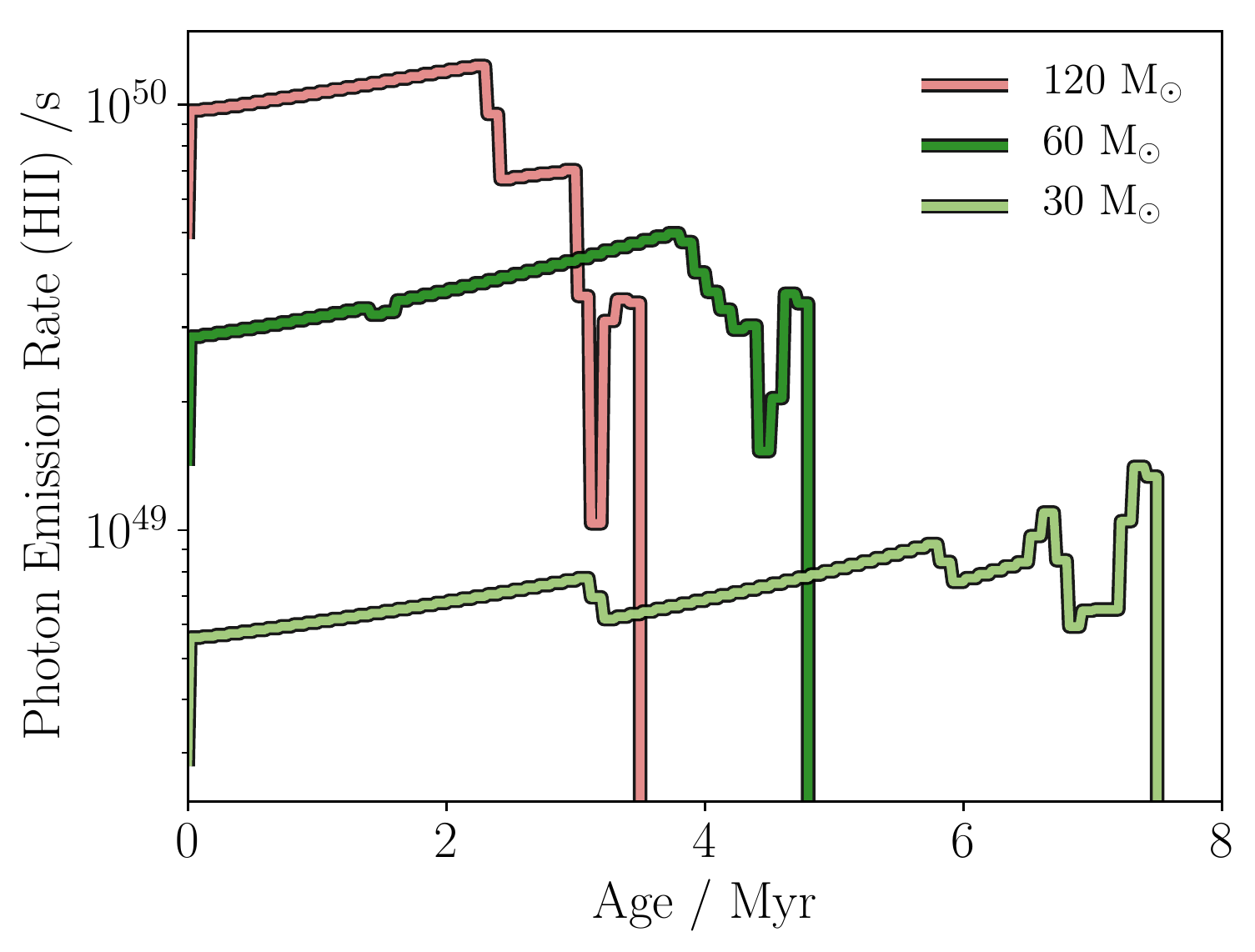}
	\includegraphics[width=1\columnwidth]{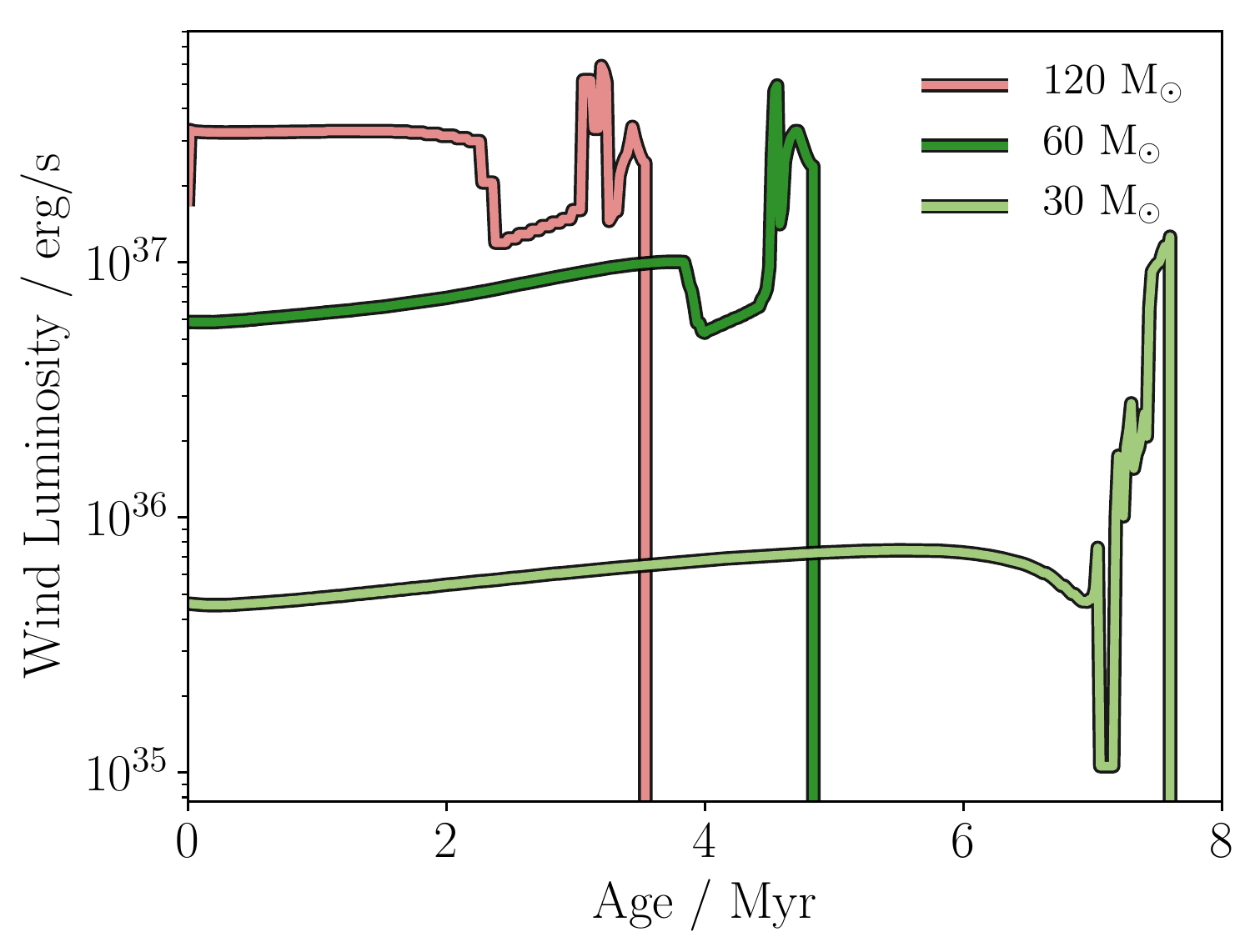}
	\includegraphics[width=1\columnwidth]{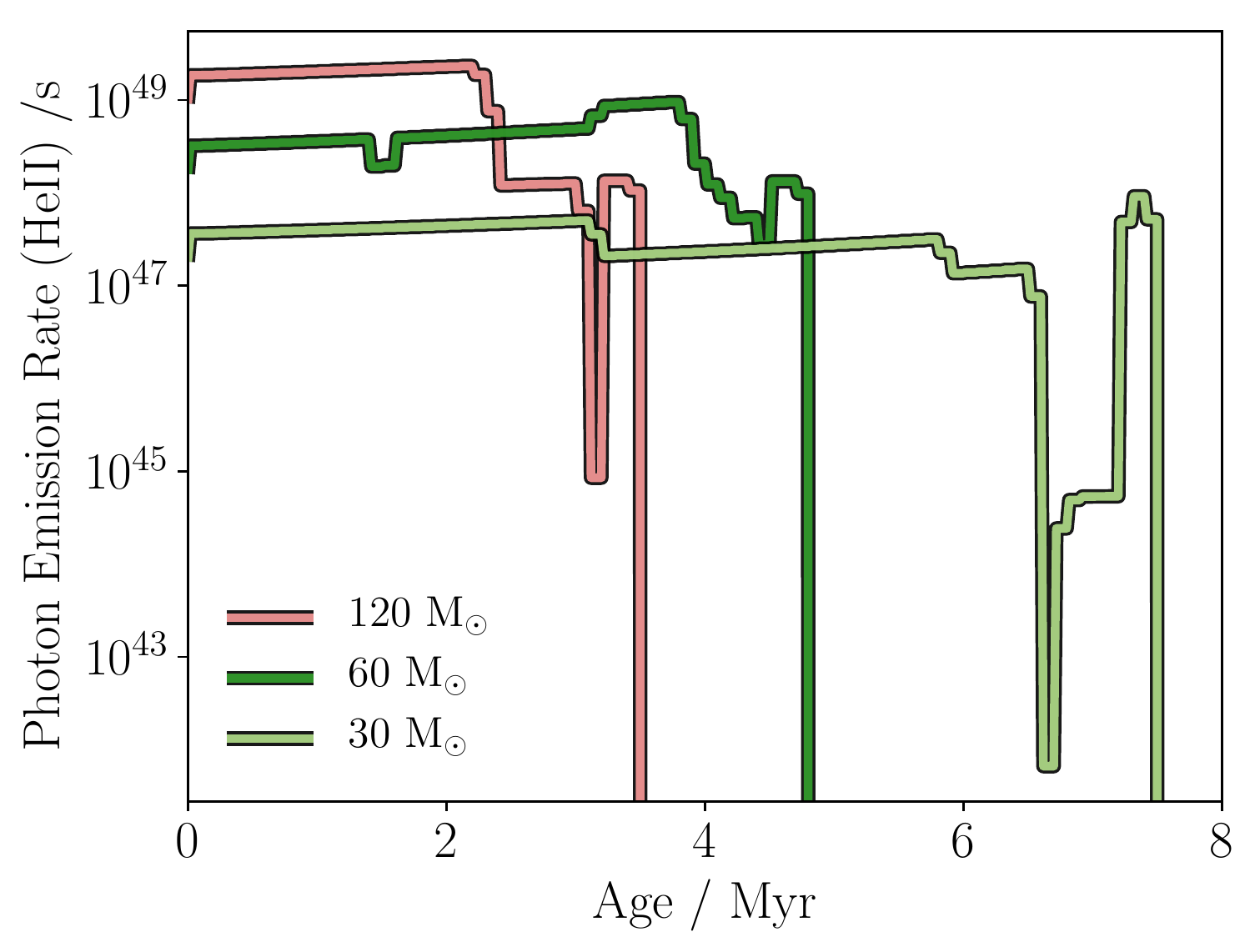}
	\includegraphics[width=1\columnwidth]{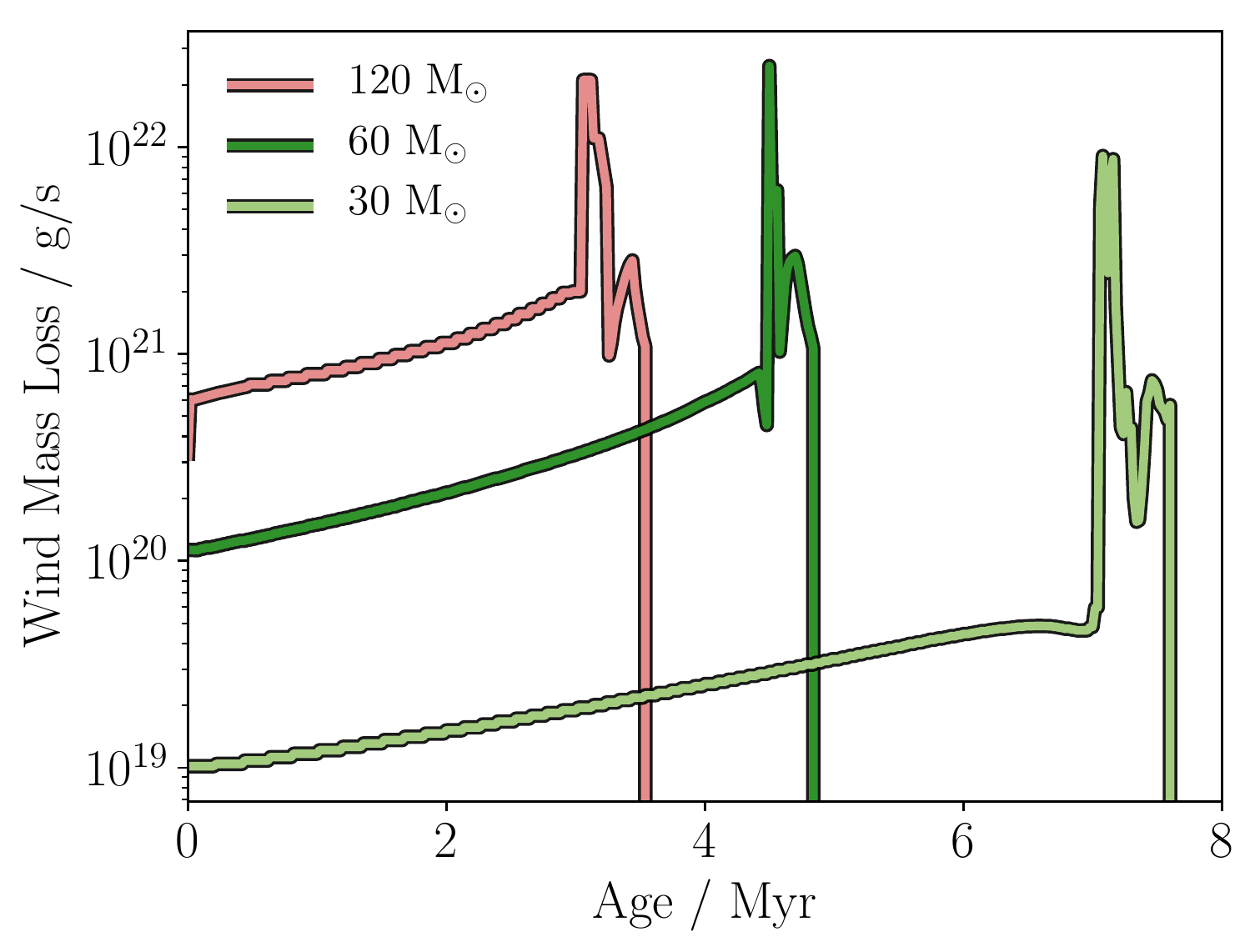}
	\includegraphics[width=1\columnwidth]{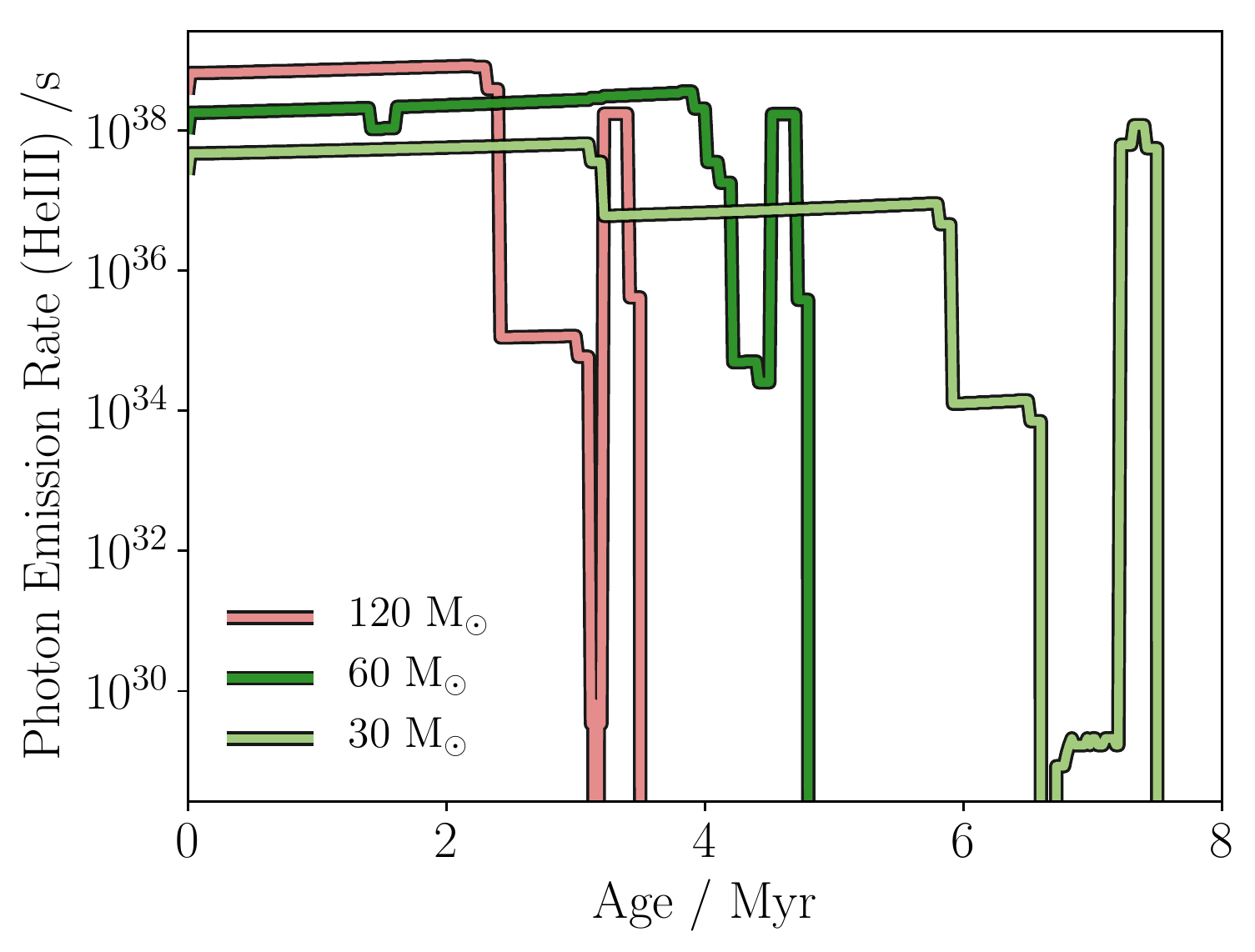}
	\caption{Wind and radiation outputs for each of the stellar tracks featured in this paper (30, 60 and 120 \Msolar stars). The left column shows wind properties: from top to bottom, cumulative energy output from winds, wind luminosity ($1/2 \dot{m}_w v_w^2$) and wind mass loss rate. The right column shows photon emission rate in groups bounded by the ionisation energy of, from top to bottom, HII, HeII and HeIII. See Section \ref{methods} for more details.}
	\label{fig:stellartracks}
\end{figure*}

In this section we plot radiation and wind properties for each of the stellar tracks featured in this paper (see Section \ref{methods}). In Figure \ref{fig:stellartracks} we plot cumulative wind energy output, wind luminosity, wind mass loss, and photon emission rates binned in photon energy bands corresponding to the ionising continua of \HI ($>~13.6~$eV), \HeI ($>~24.6~$eV) and \HeII ($>~54.2~$).

%%%%%%%%%%%%%%%%%%%%%%%%%%%%%%%%%%%%%%%%%%%%%%%%%%

% Don't change these lines
\bsp	% typesetting comment
\label{lastpage}
\end{document}